\documentclass[10pt,journal,compsoc]{IEEEtran}
\usepackage{amsmath}
\usepackage{amssymb}
\usepackage{mathabx}
\usepackage{color}
\usepackage{float}
\usepackage{graphicx}
\usepackage{multirow}
\usepackage[draft]{pgf}
\usepackage{placeins}
\usepackage{ragged2e}
\usepackage{subfiles}
\usepackage[ruled,linesnumbered]{algorithm2e}
\usepackage[caption=false,labelformat=simple]{subfig}

\usepackage{soul}

\definecolor{darkspringgreen}{rgb}{0.09, 0.45, 0.27}

\newcommand{\blue}[1]{\textcolor{blue}{#1}}
\newcommand{\red}[1]{\textcolor{red}{#1}}

\newcommand{\orange}[1]{\textcolor{orange}{#1}}

\usepackage[final]{hyperref} 
\hypersetup{
  colorlinks=true,   
  linkcolor=red,     
  citecolor=blue,    
  filecolor=magenta, 
  urlcolor=darkspringgreen
}

\newcommand{\Nk}[2]{\ensuremath{{#1}^{({#2})}}}
\DeclareMathOperator*{\argmin}{arg\,min}

\newcommand\fst[1]{\textcolor{red}{\underline{#1}}}
\newcommand\snd[1]{\textcolor{blue}{{#1}}}

\ifCLASSOPTIONcompsoc
  \usepackage[nocompress]{cite}
\else
  \usepackage{cite}
\fi
\ifCLASSINFOpdf
\else
\fi
\begin{document}
%
\title{High-dimensional Dense Residual Convolutional Neural Network for Light Field Reconstruction}
%
%
%
%

\author{Nan~Meng,~\IEEEmembership{Student Member,~IEEE,}
        Hayden~K.-H.~So,~\IEEEmembership{Senior Member,~IEEE,}
        Xing~Sun,
        and~Edmund~Y.~Lam,~\IEEEmembership{Fellow,~IEEE}

\thanks{
N.~Meng, H.K.H.~So, and E.Y.~Lam are with the Department
of Electrical and Electronic Engineering, The University of Hong Kong, Pokfulam, Hong Kong
(e-mail: nanmeng@eee.hku.hk, hso@eee.hku.hk, elam@eee.hku.hk)

X.~Sun is with the YouTu Lab, Tencent 
(e-mail: winfredsun@tencent.com)
}
}

\markboth{Journal of \LaTeX\ Class Files,~Vol.~14, No.~8, August~2015}%
{Shell \MakeLowercase{\textit{et al.}}: Bare Demo of IEEEtran.cls for Computer Society Journals}
%



\IEEEtitleabstractindextext{%
{\justify
\begin{abstract}
\label{sec:abstract}
We consider the problem of high-dimensional light field reconstruction and develop a learning-based framework for spatial and angular super-resolution. Many current approaches either require disparity clues or restore the spatial and angular details separately. Such methods have difficulties with non-Lambertian surfaces or occlusions. In contrast, we formulate light field super-resolution (LFSR) as tensor restoration and develop a learning framework based on a two-stage restoration with 4-dimensional (4D) convolution. This allows our model to learn the features capturing the geometry information encoded in multiple adjacent views. Such geometric features vary near the occlusion regions and indicate the foreground object border. To train a feasible network, we propose a novel normalization operation based on a group of views in the feature maps, design a stage-wise loss function, and develop the multi-range training strategy to further improve the performance. Evaluations are conducted on a number of light field datasets including real-world scenes, synthetic data, and microscope light fields. The proposed method achieves superior performance and less execution time comparing with other state-of-the-art schemes. 
\end{abstract}
}

}

\maketitle

\IEEEdisplaynontitleabstractindextext

%
\IEEEpeerreviewmaketitle

\IEEEraisesectionheading{\section{Introduction}\label{sec:introduction}}

\IEEEPARstart{L}{ight} field (LF) camera can capture the 3D information about an object or a scene. Compared with traditional 2D imaging systems, such cameras record the intensity of each direction of light rays passing through the lens~\cite{Ng2005Light,Lam2015Computational}. The additional information enables many applications in computer vision and imaging, such as refocusing~\cite{Mitra2012Light}, view synthesis~\cite{Srinivasan2017Learning,Kalantari2016Learning} and depth estimation~\cite{Wang2016Depth,Shin2018Epinet,Sun2016Data}.

Commercial LF cameras make use of an array of micro-lenses, placed between the main lens and the sensor, to record the spatial and angular information in a single exposure~\cite{Ng2005Light}. There is a tradeoff in resolution, such that a dense angular sampling necessarily leads to a sparse spatial sampling, and vice versa~\cite{Ng2005Light,Georgiev2010Focused}. Over the years, several approaches to achieve light field super-resolution (LFSR) have been proposed. Many of them, however, require depth estimation as a first step; that often relies on the Lambertian assumption and works poorly on glossy surfaces such as metals, plastics, or ceramics~\cite{Chan2007Super,Bishop2012Light,Mitra2012Light,Wanner2014Variational}. Occlusion also presents an additional challenge and can easily lead to artifacts in the super-resolution reconstruction.

Convolutional neural networks (CNNs) have recently been used for LFSR by learning a mapping directly from low-resolution (LR) images to high-resolution (HR) images~\cite{Yoon2015Learning,Wu2018Light,Kalantari2016Learning}. Despite delivering results generally superior to depth-based methods, several issues remain to be addressed. Chief among them is that CNNs have not been fully exploited for LF due to the complexity of the 4D data. Existing methods implement CNNs on neighboring views~\cite{Yoon2015Learning} or epipolar plane images (EPIs)~\cite{Wu2018Light}, considering only 2D information when training the network. Therefore, the features reflecting the inherent structure of LF is not fully represented and extracted. In addition, the reconstruction process is applied on individual sub-aperture or EPI images, resulting in inefficiency of such algorithms. 

To address the problems, we explore solutions from the higher order and propose a deep high-dimensional dense residual network (\textbf{HDDRNet}) to extract the representative features encoded with geometry information for LFSR. To alleviate the training of high-dimensional network, we apply the batch normalization~\cite{Ioffe2015Batch} and improve the whiten process by considering the view correlations in feature space. Our network naturally accommodates the LF data and reconstruct the entire scene progressively.
The model consists of a spatio-angular restoration network, followed by a refinement of the details. The former uses densely-connected high-dimensional residual blocks to reconstruct the light distribution information, while the latter generates visually realistic spatial details while preserving angular correlations. Instead of using the $\ell_2$ loss function to supervise the entire network, we propose to train the latter stage with the aperture-wise perceptual loss function to improve the reconstruction quality of spatial details. Although both stages contain multiple high-order operations, we are able to train the network in an end-to-end fashion without stage-wise optimization.

The main contributions of this study are:
\begin{itemize}
    \item \textbf{High-order convolution.} We incorporate high-order convolution within a deep learning architecture to super-resolve LFs, achieving reconstruction at multiple scales in spatial or angular dimensions, or both. 
    Such an approach allows the model to learn representations with scene geometry information by fully exploiting the high-dimensional LF data, enhancing the performance of synthesizing novel views.
    \item \textbf{Geometric features.} We reveal that the high-order convolution possess the potential to extract features endowed with geometry information, named geometric features. The geometric features vary near the occlusions and therefore indicate the foreground object border.
    \item \textbf{Progressive reconstruction.} Our model reconstructs the high-quality LF in one feedforward pass through two sub-networks. The first is trained by optimizing the angular loss based on mean square error (MSE), which is crucial for learning the light distribution, while the second is trained by minimizing the perceptual loss~\cite{Johnson2016Perceptual}. This achieves a more realistic spatial reconstruction while preserving the learned light distribution properties from the previous network. 
    \item \textbf{Multi-range training.}
    To train the network more effectively, we further propose a strategy of learning that exploits the spatial inter-scale correlations and multiple angular baseline range to achieve higher reconstruction accuracy. Such a multi-range model is termed \mbox{\textbf{M-HDDRNet}}.
\end{itemize}

\section{Related Work}
\label{sec:related_work}
Many LFSR approaches focus on enhancing either the spatial or the angular resolution, and accordingly we review them briefly in separate sections.

\subsection{Spatial super-resolution}
Spatial super-resolution generally makes use of sub-aperture images, in a manner similar to single-image super-resolution. However, with LF, the achievable resolution of a sub-aperture image can be beyond the limitation of the lenslet array that splits incoming light in different directions. As discussed in~\cite{Liang2015Light,Wu2017LightTIP}, the intensity values in neighboring views are propagated to the target view with non-integer shifts between two corresponding pixels. This becomes apparent when considering EPIs; as such, several methods are designed to analyze the scene geometry first, and compute pixel intensity based on the estimated disparity information.

In~\cite{Bishop2012Light}, Bishop and Favaro propose a Bayesian framework to restore more information from the geometric structure of the scene by analyzing the correlations between adjacent views. Lim et al.~\cite{Lim2009Improving} show that the angular data provide the subpixel shift information used by many SR algorithms. Wanner and Goldluecke~\cite{Wanner2014Variational} optimize a variational framework to enhance the resolution of novel views in a scene. Meanwhile, Mitra and Veeraraghavan~\cite{Mitra2012Light} propose a patch-based model based on Gaussian mixture and reconstruct the patches according to the subpixel shift. These disparity-based methods however are problematic for occlusion regions and non-Lambertian surfaces, where the estimation algorithms can fail easily and result in artifacts such as tearing and ghosting. 

Taking advantages of CNNs, some recent learning-based methods aim to be free from the disparity estimation step. Yoon et al.~\cite{Yoon2015Learning} are among the first to apply CNN-based model to perform LFSR.
However, their model treats the spatial and angular information separately, underusing the potential of the entire angular information. Considering the angular correlation, Wang et al.~\cite{Wang2018Lfnet} adopt a bidirectional recurrent CNN framework on horizontal or vertical sub-aperture images to model the spatial correlation iteratively. Meanwhile, Farrugia and Guillemot~\cite{Farrugia2018Light} apply a deep CNN on the aligned sub-aperture images to restore the entire light field. By considering light field as image sequences, these attempts to some extent exploit the subpixel shift among adjacent views. However, the light field imaging systems sample the light distribution on every spatial pixel from a 2D angular space. Such relationship is not fully represented in the image sequences,  thus limiting the performance of these methods.

\subsection{Angular super-resolution}
Angular LFSR, also commonly called view synthesis, is based on two different approaches. The first employs depth estimation algorithms~\cite{Jeon2015Accurate,Tao2013Depth,Wang2015Occlusion} to acquire an accurate depth map and then warps the existing images to the novel views~\cite{Kalantari2016Learning,Wanner2014Variational,Wanner2012Spatial}. For example, an automatic depth layer-based method is introduced in~\cite{Pearson2013Plenoptic} to generate an arbitrary view with a probabilistic interpolation approach, and depth information is calculated on a small set of sub-aperture images.
Zhang et al.~\cite{Zhang2013Light} reconstruct the LF from a micro-baseline stereo pair. They introduce a phase-based synthesis strategy to integrate disparity into the phase term when warping the input view to any close novel view, and a subsequent work further develops a patch-based synthesis method~\cite{Zhang2017Plenopatch}. However, the quality of depth estimation depends on the scene content, and as such, these methods often introduce visual artifacts in the synthesized views.

The second set of approaches formulate the view synthesis as sampling and consecutive reconstruction of the plenoptic function~\cite{Levoy1996Light}, where every pixel of the given views is considered a sample of a multidimensional LF function. Levin and Durand~\cite{Levin2010Linear} propose a linear algorithm using a dimensionality gap prior to render a LF from a 3D focal stack sequence without depth estimation. Vagharshakyan et al.~\cite{Vagharshakyan2018Light} consider the view synthesis as an inpainting task on EPI, and use the sparse representation of LF in shearlet transform to enhance the angular resolution.

Nevertheless, both set of approaches above are vulnerable to scenes with non-Lambertian surface, leading to researchers developing learning-based algorithms in recent years. Flynn et al.~\cite{Flynn2016Deepstereo} synthesize novel views based on a sequence of images with wide baselines. Kalantari et al.~\cite{Kalantari2016Learning} use two sequential CNNs to model depth and estimate color simultaneously. The disparity information and input views are then warped into the novel view. However, such depth-dependent method easily results in ghosting artifacts in the occluded regions. Gul and Gunturk~\cite{Gul2018Spatial} propose an algorithm for LFSR using two sequential CNNs. By combining the CNN models with different functions, their approach achieves both spatial and angular enhancement. Yet, with such pixel-level reconstruction strategy, the results easily suffer from jagging and lattice artifacts near the edges. Wu et al.~\cite{Wu2018Light} exploit the clear texture structure of the EPI and adopt a CNN to restore the EPI angular information. By making full use of EPI properties, the restored novel view is more pleasant compared with previous attempts. However, their network reconstructs a LF by restoring every EPI, which severely restricts the efficiency of the algorithm. Other than these EPI-wise~\cite{Wu2018Light}, pixel-wise~\cite{Gul2018Spatial} or aperture-wise~\cite{Kalantari2016Learning,Yoon2017Light} reconstruction schemes, we propose a novel schemes with throughput of the entire LF, and therefore improve the efficiency of the practical reconstruction.

\section{Problem Analysis and Formulation}
\label{sec:problem_analysis}
\subsection{Light field representation}
\begin{figure}[t]
\centering
\includegraphics[width=0.6\columnwidth]{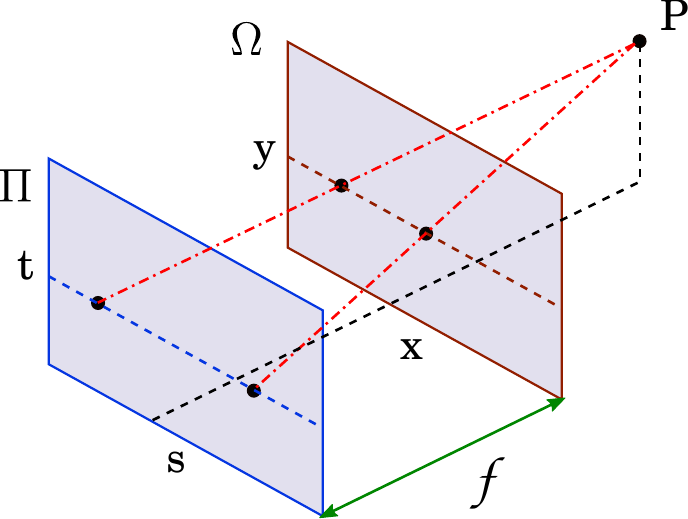}
\caption{Two-plane parameterization of light field imaging.}\label{fig:two-planes}
\end{figure}

We consider the simplified representation of light field~\cite{Levoy1996Light} or lumigraph~\cite{Gortler1996Lumigraph}, describing the propagation of light rays by a 4D function $L(x,y,s,t)$. In this representation, a light field is a collection of images captured by several cameras with the view points parallel to a common image plane, as shown in Fig~\ref{fig:two-planes}. The focal plane contains the view points which are indexed by the coordinates $(s,t)$, and the image plane is parameterized by the coordinates $(x,y)$. A 4D light field is thus a mapping
$(x,y,s,t) \rightarrow L(x,y,s,t), \quad L : \Omega \times \Pi \rightarrow \mathbb{R}.$
The mapping can be regarded as an assignment of an intensity value to each radiance of rays passing through the two planes. 

\subsection{Problem formulation}\label{sec:problem_formulation}
We treat LFSR as a high-dimensional tensor restoration. Consider a given LR light field $I^{\mathrm{LR}} \in \mathbb{R}^{X \times Y \times S \times T}$, which is equivalent to downsampling an HR light field $I^{\mathrm{HR}} \in \mathbb{R}^{r_s X \times r_s Y \times r_a S \times r_aT}$ by two factors $r_s$ and $r_a$, where $X$ and $Y$ denote the embedded spaces defined by spatial coordinates and $S$ and $T$ denote the angular embedded spaces. We use $r_s$ and $r_a$ to describe the respective scaling factors. The learning-based super-resolution can be described as
\begin{equation}
I^{\mathrm{SR}}\left(x, y, s, t\right) = g\left(I^{\mathrm{LR}}\left(x, y, s, t\right); \Theta\right),
\end{equation}
where $\Theta = \left \{\Nk{\theta}{0},\Nk{\theta}{1},\ldots,\Nk{\theta}{K-1} \right\}$ represents the parameters of the networks, and $g(\cdot)$ describes the learned mapping from LR to HR light fields. The deep learning model learns the mapping hierarchically through a stack of layers. Each layer is parameterized by a collection of weights and biases $\Nk{\theta}{k}=\left \{\Nk{W}{k},\Nk{b}{k} \right \}$, followed by a nonlinear activation function $\Nk{\delta}{k}$, where $k \in \left[ 0,K-1 \right]$. Thus, the mapping from layer $k-1$ to $k$ can be expressed as
\begin{equation}\label{equ:forward}
\Nk{g}{k}\left(I^{\mathrm{LR}}; \Nk{\theta}{k}\right) 
= \Nk{\delta}{k} \left(\Nk{W}{k} * \Nk{g}{k-1} \left(I^{\mathrm{LR}};\Nk{\theta}{k-1}\right) +\Nk{b}{k}\right),
\end{equation}
for $k \geq 1$. 

Moreover, the function $g(\cdot)$ can be considered as the composition of multiple mappings, i.e., $g = \Nk{g}{K-1} \circ \Nk{g}{K-2} \circ \ldots \circ \Nk{g}{0}$, where the symbol $\circ$ represents function composition.
The original mapping is set to be identical, such that $\Nk{g}{0}\left(I^{\mathrm{LR}};\Nk{\theta}{0}\right) = I^\mathrm{LR}$. 
All of the model parameters are optimized to reduce the loss $\mathcal{L}(\cdot)$, which measures the difference between $I^{\mathrm{SR}}$ and $I^{\mathrm{HR}}$. Thus, the light field SR problem can be formulated as 
\begin{equation}
\Theta^{*} = \underset{\Theta}{\argmin} \ \mathcal{L} \left(I^{\mathrm{HR}}, g\left(I^{\mathrm{LR}}; \Theta\right)\right).
\end{equation}
Our proposed network directly learns the mapping $g(\cdot)$ between LR light field inputs and HR labels, and reconstruct the entire light field in a single feedforward propagation.

\begin{figure}[t]
\centering
\subfloat[4D Convolution with receptive field highlighted.]{
    \includegraphics[width=0.39\columnwidth]{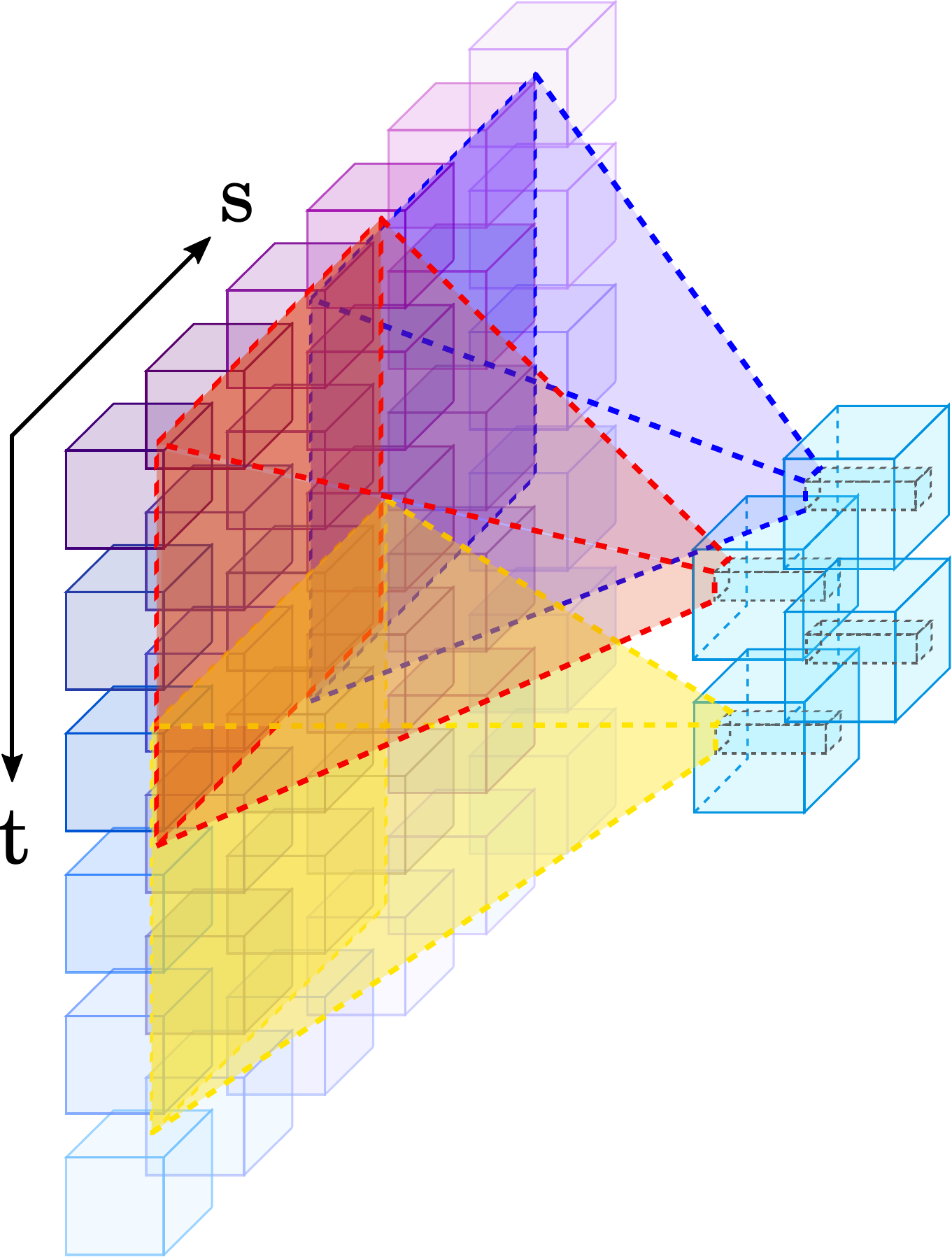}\label{fig:convolution_a}
    }\hspace{3mm}
\subfloat[The details of 4D feedforward convolutions.]{
    \includegraphics[width=0.5\columnwidth]{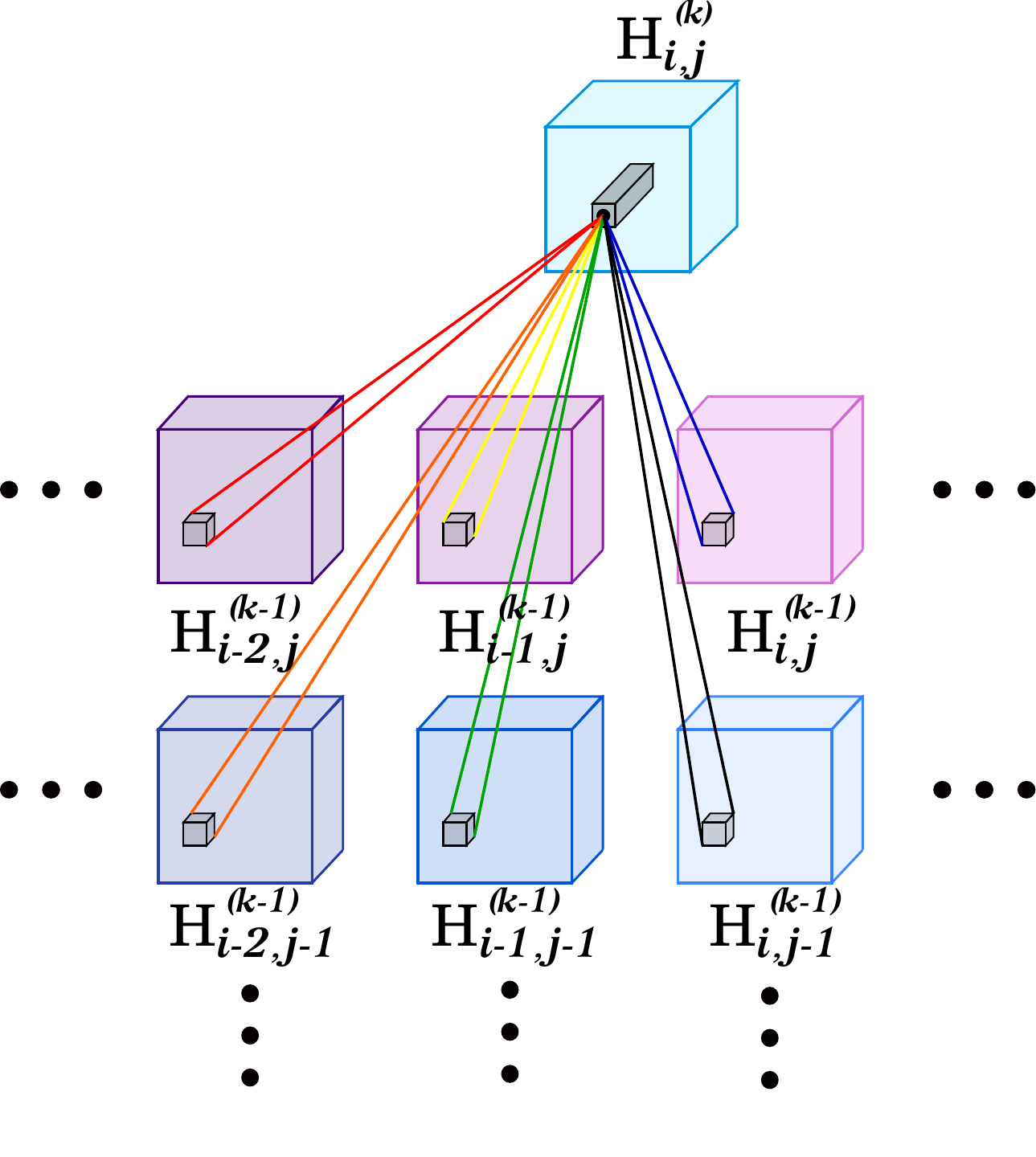}\label{fig:convolution_b}
}
\caption{The details of 4D feedforward convolutions on both spatial and angular dimensions.}\label{fig:convolution}
\end{figure}

\subsection{4D convolutional neural networks}
\label{subsec:4dconv}
Ordinarily for images, convolutions are applied on the 2D feature maps, However, for light field, it is more desirable to capture the spatio-angular information encoded in multiple adjacent views. Nevertheless, due to limitations in the traditional CNNs designed for 2D images, most existing learning-based methods apply them only on adjacent sub-aperture images~\cite{Yoon2015Learning,Kalantari2016Learning} to learn the relationships along angular coordinates, or on EPI images\cite{Wu2017Light} to model scene geometry along one  angular and one spatial coordinates. These approaches tend to underuse the potential of light field, leading to artifacts in the region with complex light distribution, such as occluded regions or non-Lambertian surfaces. By convolving a 4D kernel with a tensor formed by cascading multiple neighboring views together in the angular dimensions, the feature maps are connected to adjacent views from the previous layer, thus capturing the spatio-angular information.

We consider the input LR sub-aperture image set as $\left \{I^{\mathrm{LR}}_{s,t}\right \}$, where $s=1,2,\ldots,S$ and $t=1,2,\ldots,T$. We use subscript to denote the position of each input sub-aperture image (or feature cube), which is shown in Fig~\ref{fig:convolution_a}. Moreover, we infer the hidden layers $\mathbf{H}^{(k)}$, where $k=0,1,\ldots,K-1$, according to Eq.~\ref{equ:forward}, and therefore 
\begin{equation}\label{equ:operation}
\mathbf{H}^{(k)} = \delta \left(\mathbf{W}^{(k)} * \mathbf{H}^{(k-1)} + \mathbf{B}^{(k)}\right).
\end{equation}
$\mathbf{W}^{(k)}$ and $\mathbf{B}^{(k)}$ represent the filters and bias of 4D feedforward convolution, respectively. 
Both have size $s_1 \times s_2 \times a_1 \times a_2 \times n$, where $n$ is the number of filters, $s_1 \times s_2$ is the spatial filter size, and $a_1 \times a_2$ is the angular filter size. To avoid the dying neuron problem in rectified linear units, we apply the leaky rectified linear units (LeakyReLU) proposed by Maas et al.~\cite{Maas2013Rectifier} as the activation function in each layer, i.e.,
\begin{equation}
\delta^{(k)}(x) = \delta(x) = \left\{\begin{matrix}
x & \text{if}\ x\geq 0\\ 
\alpha x & \text{if}\ x < 0
\end{matrix}\right. .
\end{equation}
In all experiments, we set $\alpha=0.2$. 
The notation $*$ in Eq.~\ref{equ:operation} is implemented using cross-correlation combining the input feature map with the filter, i.e.
\begin{equation}
\begin{aligned}
h_j^{(k)}(x,y,s,t) = &\sum_{i=0}^{c-1}\sum_{m=0}^{s_1-1}\sum_{n=0}^{s_2-1}\sum_{u=0}^{a_1-1}\sum_{v=0}^{a_2-1}w_{i,j}^{(k)}(m,n,u,v) \cdot \\
&h_i^{(k-1)}(x+m,y+n,s+u,t+v),
\end{aligned}
\end{equation}
where $h_j^{(k)}(x,y,s,t)$ is the value at position $(x,y,s,t)$ on the $j^{\mathrm{th}}$ feature map $h_j^{(k)}$ in $\mathbf{H}^{(k)}$, and $w_{i,j}^{(k)}(m,n,u,v)$ is the value at position $(m,n,u,v)$ in the filter connected between the $i^{\mathrm{th}}$ stacked input channel and the $j^{\mathrm{th}}$ feature map.

\begin{figure}[t]
\centering
\subfloat[2D feature slices]{
    \includegraphics[width=0.48\columnwidth]{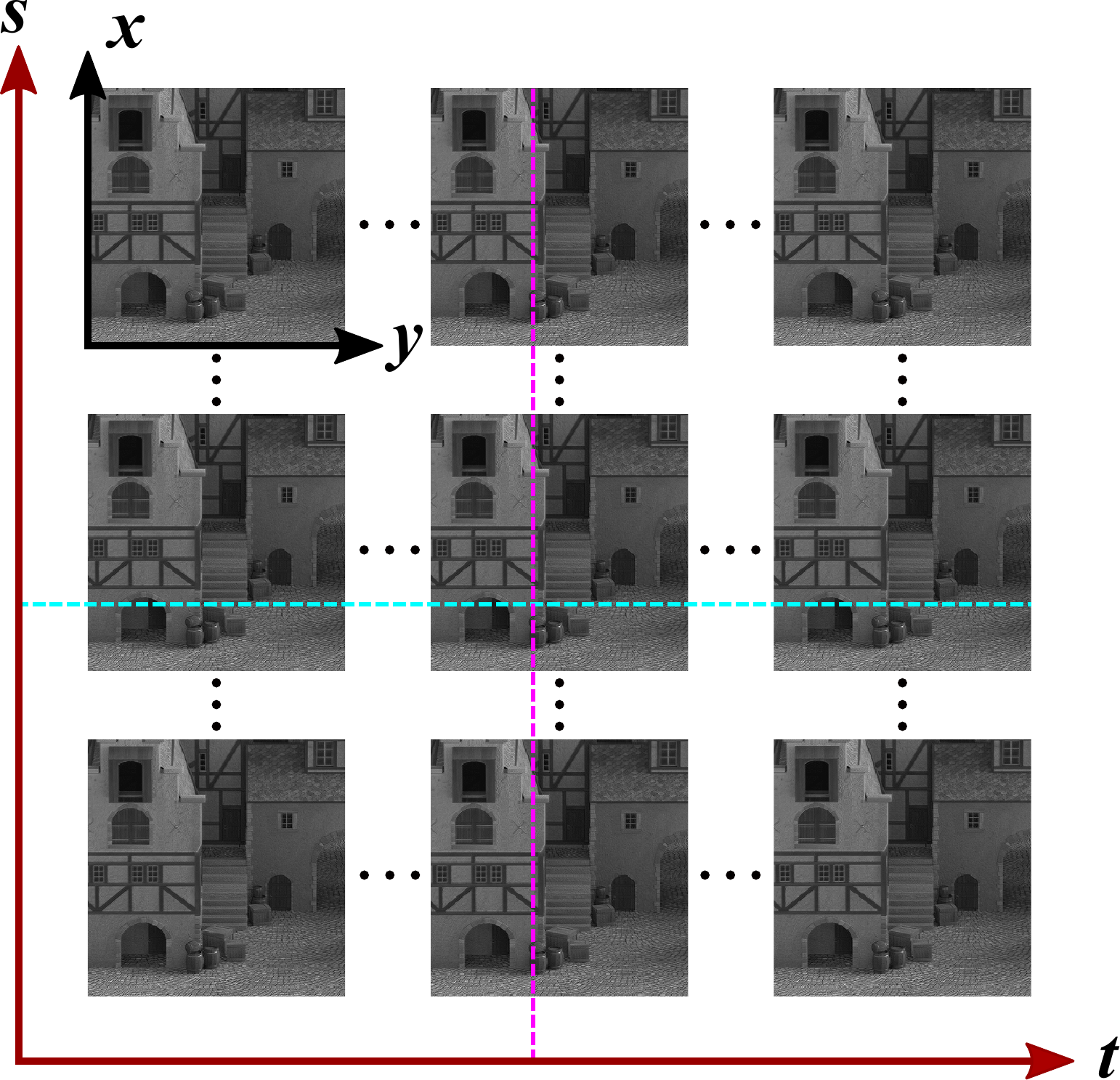}\label{fig:geometric_features_a}
    }
\subfloat[Feature EPIs]{
    \includegraphics[width=0.48\columnwidth]{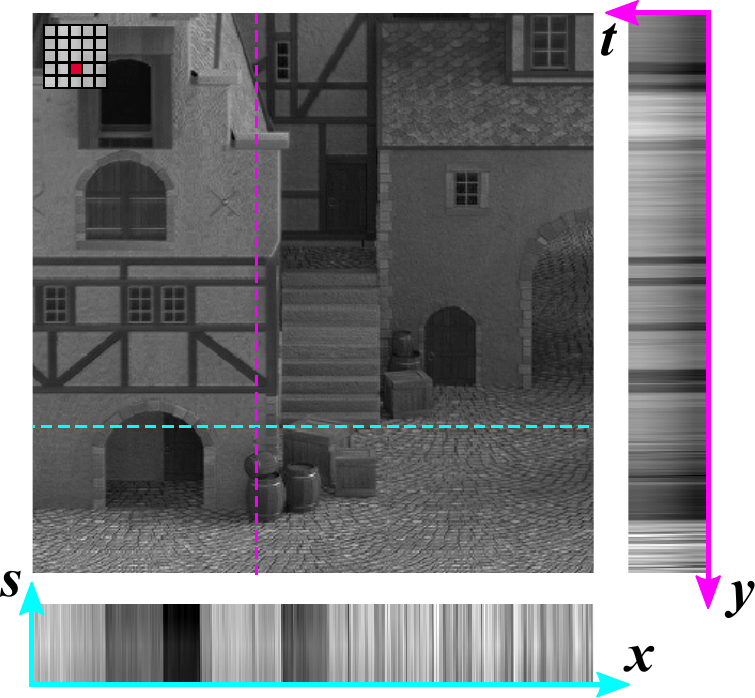}\label{fig:geometric_features_b}
    } \\
\subfloat[Reconstruction]{
    \includegraphics[width=0.48\columnwidth]{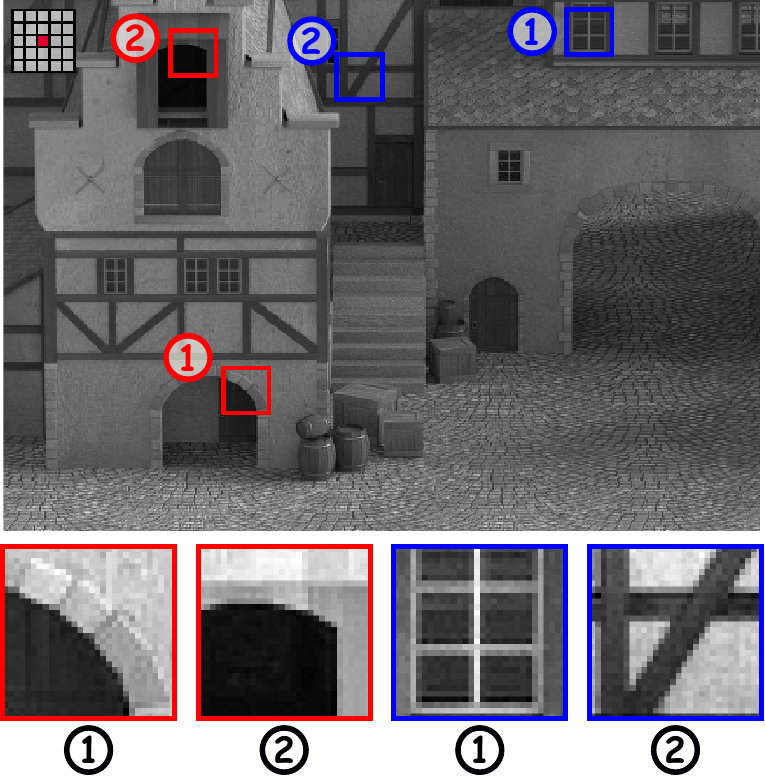}\label{fig:geometric_features_c}
    }
\subfloat[Feature (Conv4D 22, Channel 4)]{
    \includegraphics[width=0.48\columnwidth]{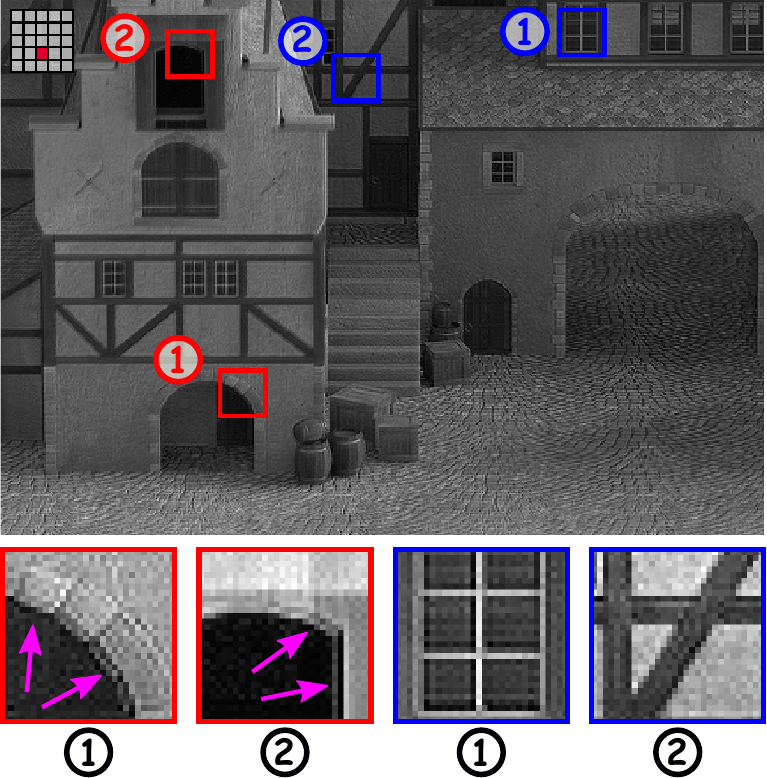}\label{fig:geometric_features_d}
    } \\
\subfloat[Feature (Conv4D 15, Channel 1)]{
    \includegraphics[width=0.48\columnwidth]{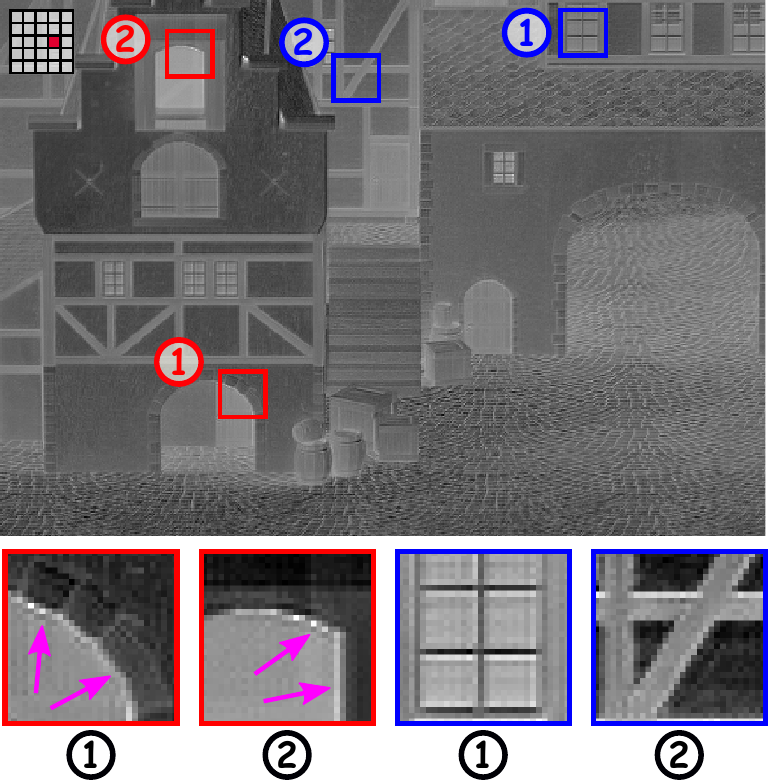}\label{fig:geometric_features_e}
    }
\subfloat[Feature (Conv4D 7, Channel 50)]{
    \includegraphics[width=0.48\columnwidth]{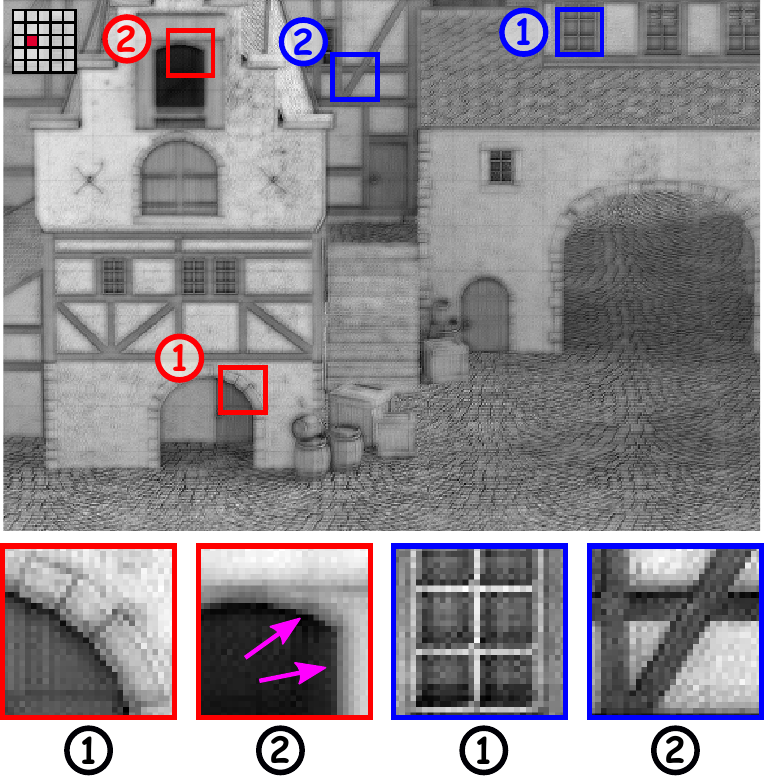}\label{fig:geometric_features_f}
    }
\caption{Visualization of the geometric features extracted from the proposed 4D framework. \textbf{(a)} The collection of 2D slices through the learned feature maps. \textbf{(b)} A certain 2D slice of the 4D geometric feature map shown in (a), and the EPIs located at corresponding lines. \textbf{(c)} The spatial reconstruction results. \textbf{(d)--(f)} geometric features extracted from different 4D convolutional layer.}\label{fig:geometric_features}
\end{figure}

\subsection{Geometric features}
\label{subsec:geometric_features}

The major benefit of using 4D convolution for light field processing is that it is able to extract the spatial features that preserve geometrical properties. Such feature maps not only contain spatial structures (e.g. textures and edges at different directions) but record the relationship of adjacent views as well. Fig.~\ref{fig:geometric_features} exhibits an example of the feature maps learned by a single 4D convolutional layer in the network. To demonstrate these high-dimensional features, we present the 2D slices through the 4D features and arrange them in an equally spaced rectangular grid in Fig.~\ref{fig:geometric_features_a}. Meanwhile, Fig.~\ref{fig:geometric_features_b} shows a certain single view and the horizontal and vertical ``feature EPIs'' acquired by gathering the feature samples with a fixed spatial coordinate and an angular coordinate. The feature EPIs are very similar to the light field EPIs, reflecting that the features learned by the 4D convolution layer have high coherence.
In addition, the geometry properties are also reflected in the spatial dimensions. The learned spatial features are sensitive to the regions with occlusions, such as the foreground object border. In Fig.~\ref{fig:geometric_features_d}, Fig.~\ref{fig:geometric_features_e} and Fig.~\ref{fig:geometric_features_f}, we visualize and compare two types of spatial features extracted from different 4D convolutional layers, namely the \emph{object border} and \emph{texture} features. The object border features are always with occlusions and displayed in the red boxes, while the texture features are presented in blue boxes. As is shown in the figures, the edges of object border features are smoother compared with the corresponding ones of reconstruction results in Fig.~\ref{fig:geometric_features_c}. By contrast, however, the edges of texture features remain clear.

\begin{figure}[t]
    \centering
    \includegraphics[width=0.8\columnwidth]{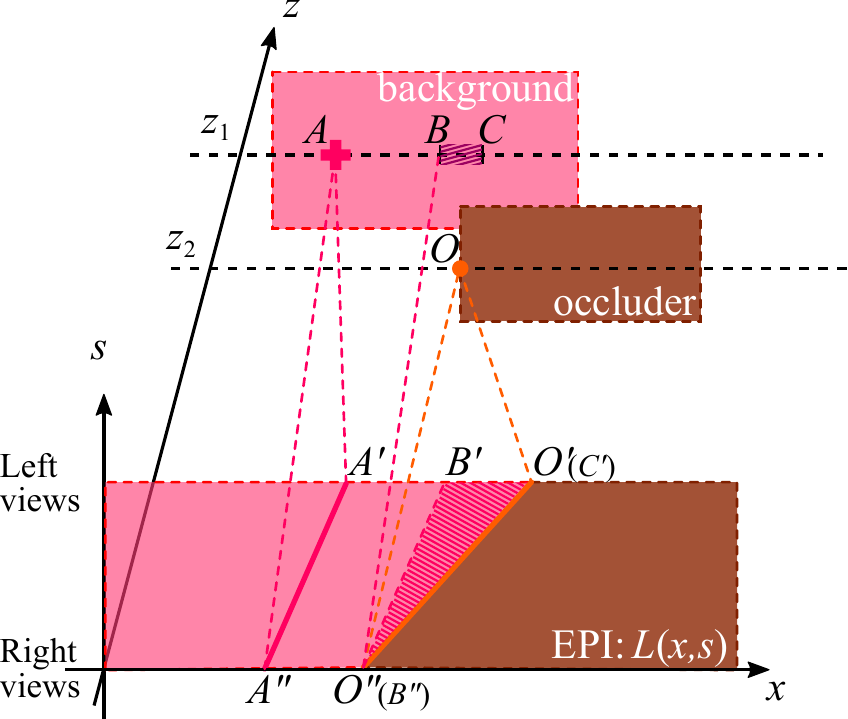}
    \caption{\textbf{Illustration of partially occluded region in EPI pattern.} The positive direction of $s$ denotes the left views.}
    \label{fig:light_model}
\end{figure}

Such smoothing effects near object border is related to the scene geometry, other than a random occurrence. To demonstrate typical variances, we analyze the light ray transmission in the LF imaging system.
Fig.~\ref{fig:light_model} exhibits an example configuration for two objects placed at different distances from the camera and the corresponding EPI pattern. The near object (denoted as ``occluder'') whose distance is $z_2$ partially occludes the further object in red (denoted as ``background'') with the distance $z_1$. We denote the positive direction of $s$ as the left views in a LF. The line $A'A''$ on EPI is projected from $A$ of the background, where point $A'$ corresponds to the leftmost view and $A''$ corresponds to the rightmost view (the same for points $B$, $O$, and $C$). The shaded region $BC$ of the background is partially occluded by the occluder at $z_2$ with no occlusion from the leftmost view and completely occluded from the rightmost view, and the corresponding region $B'O'O''$ on the EPI is defined as partially occluded region (POR). As a result, in the POR, the pixels belong to occluder shifts with larger distance than the background pixels among different views (in both the input and the feature space). Considering the 4D convolutional layer is approximately linear (the LeakyReLU is piecewise linear), the features of each layer are actually calculated as a weighted combination of multiple views directly or indirectly from inputs. The features near object border is therefore smooth.

\subsection{Aperture group batch normalization}
To ease the training of 4D framework, we follow the work~\cite{Ioffe2015Batch} and apply the normalization to the outputs of every 4D convolutional layer. However, as is illustrated in Section~\ref{subsec:geometric_features}, considering such geometric features preserve the high coherence among adjacent views, the whiten process should not be applied on every view of the feature maps. Therefore, we implement the normalization transform over a group of sub-aperture images in each channel of the feature maps, and named the proposed operation as \textbf{aperture group batch normalization} (AGBN).

Following the description is Section~\ref{subsec:4dconv}, we consider the output of a particular hidden layer $\mathbf{H}$ (omit the superscript $k$ for brevity). We only count on the angular dimension and use a new symbol to denote the learned features in an aperture-wise manner as $\mathbf{H} = \{\mathcal{H}^p_i(s,t)\}$, where $s=1,2,\cdots,S$ and $t=1,2,\cdots,T$ are two indices of angular dimensions, and $p$ denotes the number of feature channels, and for each sub-aperture feature map contains $m$ values ($i=1,2,\cdots,m$). Then, the algorithm can be described in the  Algorithm~\ref{algo:AGBN}.
\begin{algorithm}[!h]
\SetKwInOut{Input}{Input}
\SetKwInOut{Output}{Output}
\Input{The features of a particular hidden layer: $\{\mathcal{H}^p(s,t)\}$;\\ Parameters to be learned $\gamma$, $\beta$}
\Output{The normalized features: $\mathcal{\hat{H}}^p(s,t)$}
\BlankLine
Initialize the $\epsilon = 0.001$;\
\BlankLine
\For{$p = 1, \cdots, N$}{
    \BlankLine
    $\mu_p \leftarrow \frac{1}{m}\sum_{i=1}^{m}(\frac{1}{ST}\sum_{s=1}^S\sum_{t=1}^T\mathcal{H}^p_i(s,t))=$\
    \BlankLine
    \hspace{9mm} $\frac{1}{mST}\sum_{i=1}^{m}\sum_{s=1}^S\sum_{t=1}^T\mathcal{H}^p_i(s,t)$;\
    \BlankLine
    $\sigma_p \leftarrow \frac{1}{mST}\sum_{i=1}^{m}\sum_{s=1}^S\sum_{t=1}^T(\mathcal{H}^p_i(s,t)-\mu_p)^2$;\
    \BlankLine
    $\mathcal{\hat{H}}^p(s,t) \leftarrow \gamma \cdot \frac{\mathcal{H}^p(s,t) - \mu_p}{\sqrt{\sigma_p^2+\epsilon}} + \beta$
}
\caption{Aperture group batch normalization}\label{algo:AGBN}
\end{algorithm}

\section{Method}
\label{sec:method}
\begin{figure*}[!ht]
\centering
\includegraphics[width=0.9\textwidth]{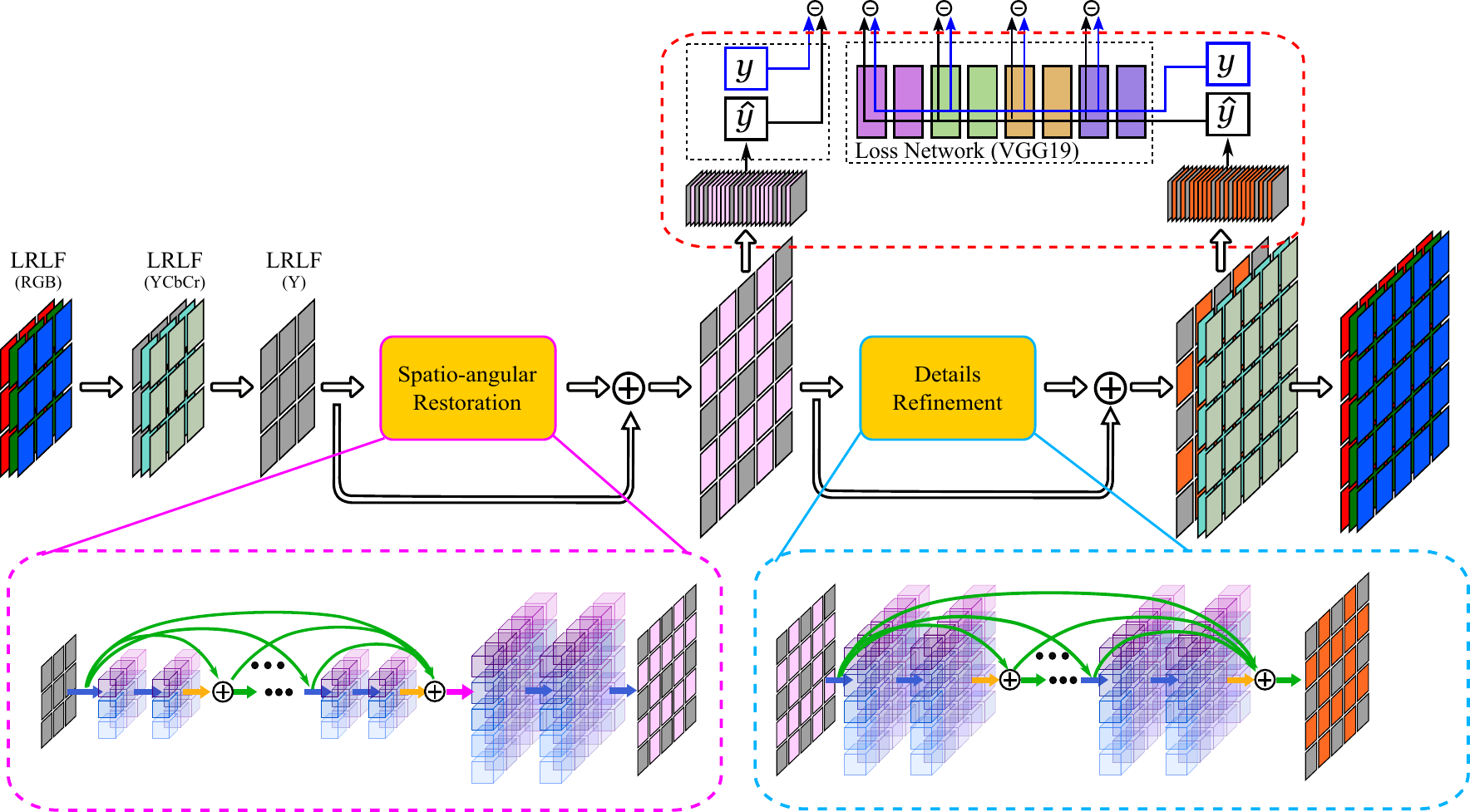}
\caption{\textbf{The overview of the proposed model.} Our model consists of a residual network for restoring the local spatio-angular information of light field and a refinement network for reconstructing the spatial details of scenes. Blue arrows indicate high-dimensional convolution operation, while yellow arrow stands for activation operation. Green arrows (with $\oplus$) indicate addition and red arrow denotes upsampling, and $\ominus$ denotes the $\ell_2$ difference.}\label{fig:LFHDRN}
\end{figure*}

\subsection{High-dimensional dense residual CNNs}
Our model is designed on the basis of the 4D convolutional layer. The network takes an LR light field as input (rather than its upscaled version) and recovers the spatial and angular information progressively. There are two subnetworks, which reconstruct the entire light field in two different stages: (1) spatio-angular restoration, and (2) details refinement.

\subsubsection{Spatio-angular restoration}
As illustrated in Fig.~\ref{fig:LFHDRN}, the spatio-angular restoration stage is set up to take down-sampled light field patches as inputs and predict the missing information. At this stage, the high-dimensional subnetwork is trained to learn the light distribution, which can assist in further super-resolving the light field. To achieve this, the network learning proceeds by minimizing the angular loss between the predicted HR light field and the ground truth using mean square error (MSE). 

\begin{figure}[t]
\centering
\includegraphics[width=1.\columnwidth]{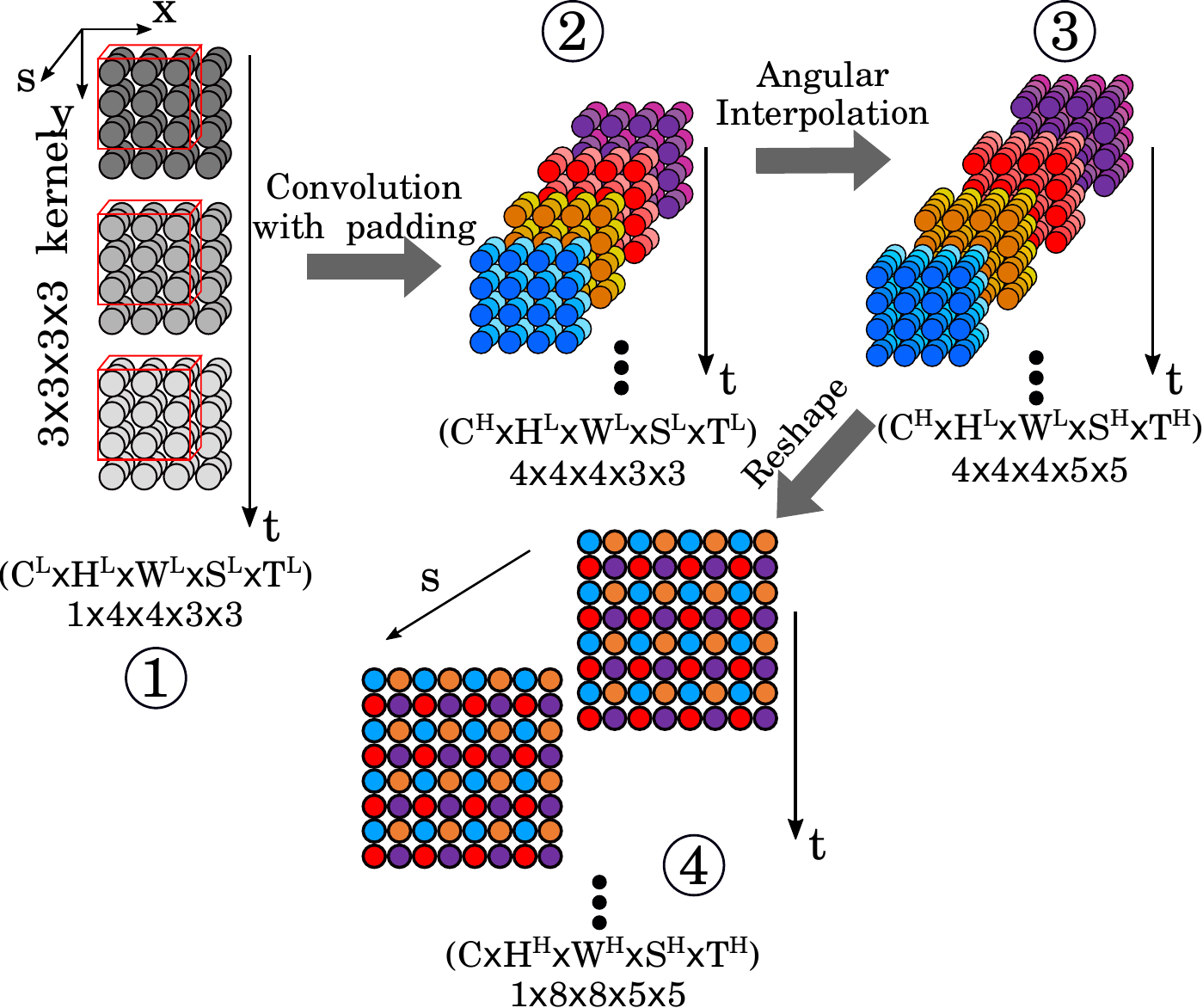}
\caption{\textbf{Upscaling operation used for resolution enhancement.} For clarity, in this example, we only consider a single feature tensor (batch size is $1$) with a single channel $C=1$. Given the LR input feature tensor with dimension $C \times H \times W \times S \times T$ ($= 1 \times 4 \times 4 \times 3 \times 3$), we first add $2$ zero-padding frames, and then apply the 4D convolution on the feature tensor. We use four 4D convolution kernels to generate the LR feature map with $4$ channels (denoted by $4$ main colors in step {\textcircled{\scriptsize $2$}}).
Subsequently, interpolation is first performed on ($S \times T$) angular dimensions of LR feature map. For spatial resolution, we applied the shuffle operation which enhances the ($H \times W$) spatial resolution of feature map and reduces the channel resolution. Therefore, at the end we have a super-resolved feature tensor of size $1 \times 8 \times 8 \times 5 \times 5$.}
\label{fig:upscaling}
\end{figure}

For upsampling, we extend the sub-pixel convolution operation proposed in~\cite{Shi2016Real} by combining it with angular interpolation to upscale an input LR feature tensor in all dimensions. A graphical illustration of the upsampling operation is presented in Fig.~\ref{fig:upscaling}. 
As an example, assuming a single channel, and the LR feature map has dimensions $H\times W\times S\times T$, where $H=W=4$ and $S=T=3$. Let the spatial upscaling factor $r_s$ and the angular upscaling factor $r_a$ both be $2$. The first step involves expanding the channel by a factor of $r_s^2$. In the second step, given the high coherence of the spatio-angular features, we use linear interpolation on the angular dimensions of the feature maps to upsample the resolution of the angular dimensions by a factor of $r_a$ each (strictly, from $3 \times 3$ to $5 \times 5$). Third, the channel-to-space transpose layer is placed on top of the feature maps to upscale both spatial dimensions by a factor of $r_s$ each.

\subsubsection{Details refinement}
The spatio-angular restoration network is trained in a supervised manner, using a mean-squared reconstruction loss to measure the difference between the output and the ground truth. While such per-pixel loss function contributes to the network learning of the angular coherence, it can also lead to difficulty in restoring the high-frequency texture. To mitigate this problem, the details refinement network, designed for recovering the spatial high-frequency details, is trained by optimizing  the sub-aperture perceptual loss. As will be demonstrated later in our experiments, such loss based on differences between high-level features is effective to drive the high-dimensional network to recover spatial details with sub-pixel accuracy.

\subsection{Loss Function}
\label{sec:loss_function}
Most  learning-based methods for light field reconstruction  use the mean squared error (MSE) between the recovered EPI image~\cite{Wu2017Light} or sub-aperture image~\cite{Farrugia2018Light,Yoon2015Learning,Wang2018Lfnet} and the ground truth. However, typical loss function encourages the model to find pixel-wise averages of plausible solutions that are often too smooth~\cite{Ledig2016Photo,Gupta2011Modified}, resulting in edge artifacts such as blurring and ghosting in the region containing complex occlusions or textures. To reconstruct realistic spatial texture details while preserving the geometric properties, we design a novel loss function that evaluates the results concerning the entire light field characteristics. The loss function used for training our proposed network is formulated as the weighted sum of an angular loss $\ell_A$ and a spatial perceptual loss $\ell_S$, i.e.,
\begin{equation}
\ell_{SA} = \alpha \cdot \ell_S + \beta \cdot \ell_A,
\end{equation}\label{equ:saloss}
where scalars $\alpha$ and $\beta$ denote the weights of each loss.

\textbf{Spatial loss} measures the quality of reconstructed light field in terms of spatial coordinates. Inspired by~\cite{Ledig2016Photo} and~\cite{Johnson2016Perceptual}, we extend the perceptual loss to describe aperture-wise differences between high-level feature representations. Such loss obtained from pre-trained 19-layer VGG network~\cite{Simonyan2014Very} encourages the network to restore the spatial information with better high-frequency details. In our experiments, the spatial loss is obtained by calculating the average value of content loss through all the sub-aperture images which can be formulated as
\begin{equation}
\ell_S = \frac{1}{ST}\sum_{s=1}^S\sum_{t=1}^T\left(f(I^\mathrm{HR}_{s,t}) - f(g(I^\mathrm{LR}_{s,t}; \Theta))\right)^2,
\end{equation}
where $f(\cdot)$ indicates the summation of all the feature maps after every activation function of VGG network. We use $I^{\mathrm{LR}}_{s,t}=I^{\mathrm{LR}}(\cdot,\cdot,s,t)$ and $I^{\mathrm{HR}}_{s,t}=I^{\mathrm{HR}}(\cdot,\cdot,s,t)$ to represent the LR input and label sub-aperture image with angular coordinates $(s, t)$, respectively. The function $g(\cdot)$ is the mapping as indicated in Section~\ref{sec:problem_formulation}.

\textbf{Angular loss} is defined on the basis of MSE between the reconstructed light field and the ground truth. This item is straightforward but critical for learning the light field structure properties. Unlike single image super-resolution, for LFSR the MSE loss not only describes the pixel-wise differences but also ensures that the results preserve the relationship of adjacent viewpoints. Such property can be reflected by rearranging the order of summation
\begin{equation}
\begin{aligned}
\ell_A &= \sum_{x=1}^X\sum_{y=1}^Y\sum_{s=1}^S\sum_{t=1}^T\left( I^{\mathrm{HR}}(x,y,s,t) - I^{\mathrm{SR}}(x,y,s,t) \right)^2\\
    &= \sum_{y=1}^Y\sum_{t=1}^T \left (\sum_{x=1}^X\sum_{s=1}^S\left( I^{\mathrm{HR}}(x,y,s,t) - I^{\mathrm{SR}}(x,y,s,t) \right)^2 \right)\\
    &= \sum_{y=1}^Y\sum_{t=1}^T\left(E^{\mathrm{HR}}(y,t) - E^{\mathrm{SR}}(y,t)\right)^2,
\end{aligned}
\end{equation}
where $E^{\mathrm{HR}}(y,t)$ and $E^{\mathrm{SR}}(y,t)$ represent the original and super-resolved EPIs acquired by gathering the light field samples in terms of a spatial coordinate $x$ and an angular coordinate $s$, respectively.

\subsubsection{Network Settings}
In the proposed HDDRNet, all 4D convolution layers have 64 filters with a spatial dimension of $3 \times 3$ and an angular dimension of $5 \times 5$. The convolution filters are initialized using the method of Glorot and Bengio~\cite{Glorot2010Understanding}. Furthermore, we use the residual blocks layout proposed by Gross and Wilber~\cite{Gross2016Training}. Each block consists of two 4D convolutional layers followed by batch normalization and the LeakyReLU~\cite{Maas2013Rectifier} with a slope $\alpha = 0.2$ in the negative domain as the non-linear activation function. 

\subsubsection{Multi-range training}
\begin{figure}[t]
\centering
\subfloat[]{
    \raisebox{3mm}{\includegraphics[width=0.50\columnwidth]{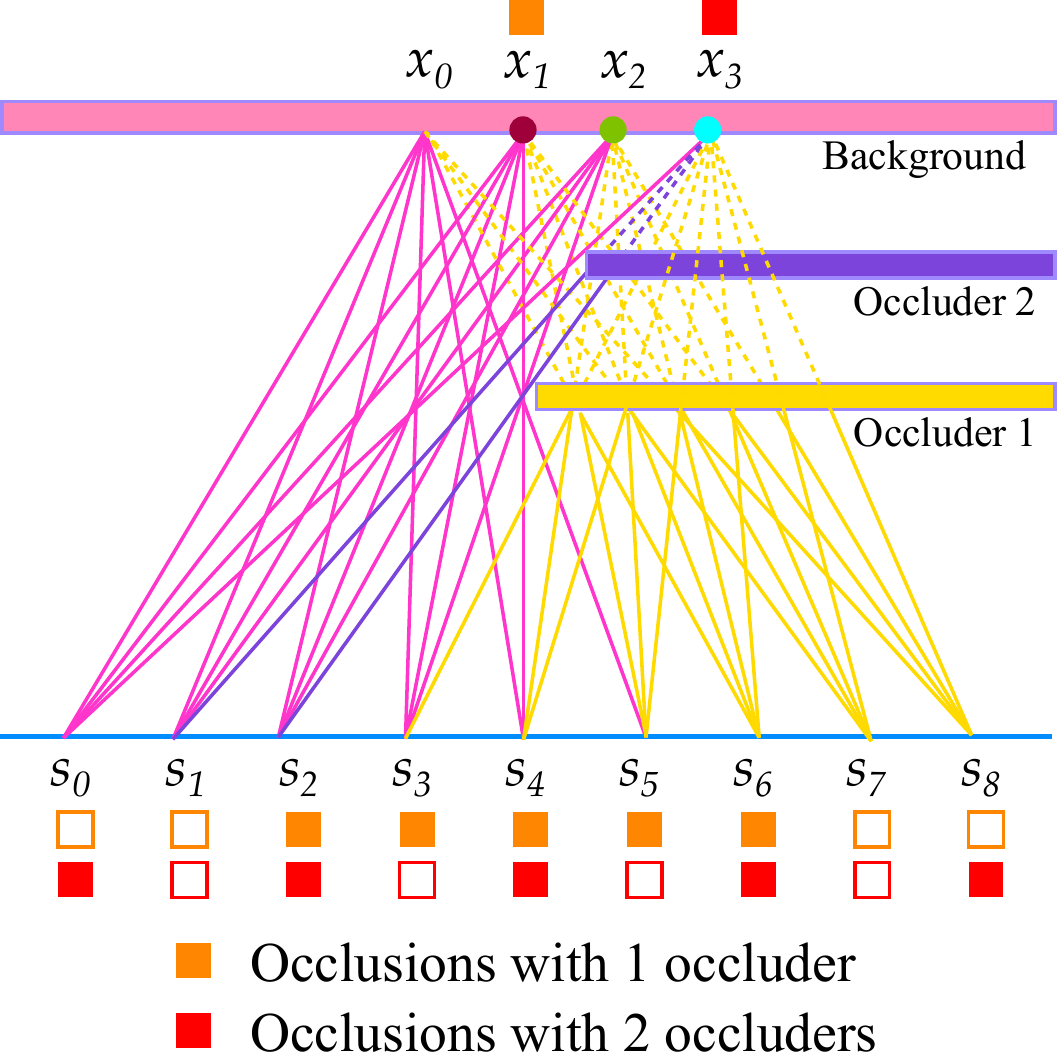}}\label{fig:multi_range_a}
    }
\subfloat[]{
    \includegraphics[width=0.42\columnwidth]{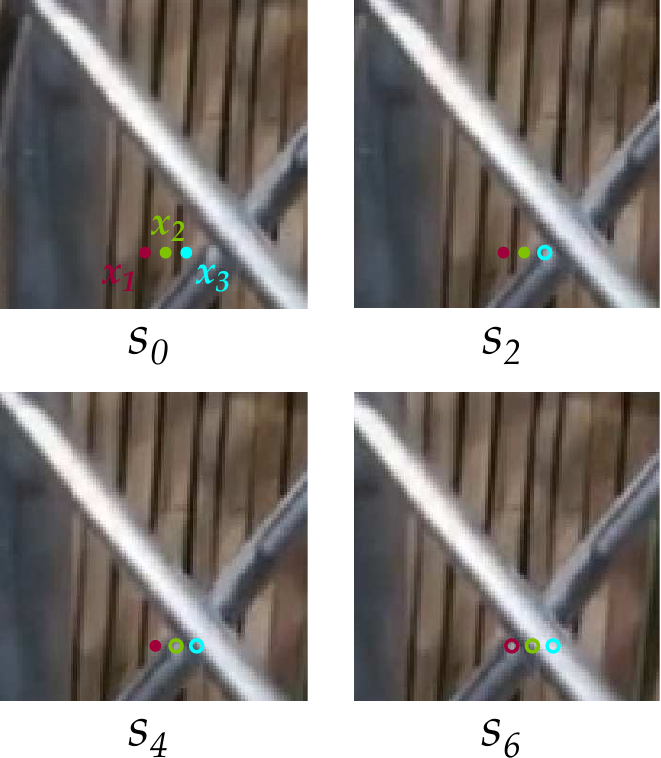}\label{fig:multi_range_b}
    }
\caption{Illustration of light distribution at the place with two occluders (a) the light ray model near occluders. The blue line denotes camera plane and $x_i$ ($i=0,1,2,3$) is a point in the background, while $s_i$ ($i=0,1,\cdots,8$) stands for the viewpoint. The orange square \orange{$\blacksquare$} denotes the selected viewpoints and pixel in background when occlusions are contributed by only 1 occluder, while the red square \red{$\blacksquare$} is used for places where the occlusions are contributed by 2 occluders. (b) illustration of light ray model in the spatial dimensions. The solid point represents the pixel without occlusion while hollow point stands for the occluded pixel.}\label{fig:multi_range}
\end{figure}

The multi-range training strategy is specifically designed for our model to learn the light distribution where there may be complex occlusions, usually at the edges of occluders. There are two major aspects to this: 1) For spatial dimension, we randomly downsample the spatial resolution between $[0.8, 1.0]$ to encourage the model to learn the inter-scale correlations~\cite{Lai2017Fast}. 2) For angular dimension, we sample 5 different angular directions with various ranges. We consider the light distribution model near the occluders as shown in Fig.~\ref{fig:multi_range_a}, where we use different colors to demonstrate the light rays from different views with occlusion. The sampling is implemented by choosing first, at random, a center view and a range, and then the surrounding views according to the range. For instance, considering the occlusions near pixel $x_1$. If one takes $s_4$ as the center view and samples the other views with range $1$, then $s_2$ to $s_6$ are selected to describe the light distribution occlusions contributed by a single occluder. If one considers the light distribution near pixel $x_3$ and takes $s_4$ as center view and samples the other views with range $2$, then $s_k$, where $k=0,2,4,6,8$, are selected to describe the light distribution with complex occlusions contributed by two occluders. An example of what the training samples look like is provided in Fig.~\ref{fig:multi_range_b}. In our experiments, the model trained using multi-range strategy has more robustness over the complex light distribution and different scaling on spatial details. Therefore, we name it \textbf{M-HDDRNet} and present the quantitative and visual comparisons in Section~\ref{sec:results}.

\subsubsection{Implementation and training details}
Our network each time receives a 4D patch of light field as the input and outputs a super-resolved 4D patch. We assume that the input LR light field patch is related to its HR counterpart based on the classical imaging model~\cite{Farrugia2017Super,Farrugia2018Light,Cheng2019Light,Rossi2018Geometry}
\begin{equation}\label{equ:downsampling}
I^{\mathrm{LR}} = \kappa(\mathrm{B} * I^{\mathrm{HR}}) + \xi,
\end{equation}
where $\xi$ represents an additive noise, $\kappa(\cdot)$ is the nearest neighbor downsampling operator on every sub-aperture image. $\mathrm{B}$ is the Gaussian kernel with window size of $7$ and standard deviation of $1.2$. The HR patches are randomly cropped from Lytro Archive~\cite{StanfordLytro} and Fraunhofer~\cite{Ziegler2017Acquisition} dataset with $96 \times 96$ pixels and $5 \times 5$ angular directions. Our model is implemented using Tensorflow toolbox~\cite{Abadi2016Tensorflow} and trained using the Stochastic Gradient Descent solver. The learning rate is initialized to $10^{-5}$ and decreased by a factor of $0.1$ for every $10$ epochs. Our implementation is available at \url{https://github.com/monaen/LightFieldReconstruction}.

\section{Experiments}
\label{sec:experiments}
\subsection{Training data and analysis}
The light fields involved in the experiments reported in this paper are all from publicly available datasets. We select 100 light fields from the Lytro Archive~\cite{StanfordLytro} (excluding occlusions and reflective) and the entire densely-sampled Fraunhofer dataset~\cite{Ziegler2017Acquisition} for training.
The former contains 353 real-world scenes captured using a Lytro Illum camera with a small baseline. Since many corner angular samples are outside the camera's aperture, for each scene, we select the center $9\times9$ views in the experiments. The latter includes 9 scenes that are densely sampled using a high-resolution camera. Each light field is processed as a $21 \times 101$ array of views, and each view is a sampling of the real-world object with a resolution as high as $1988 \times 1326$ pixels. The Fraunhofer dataset enables the proposed multi-range strategy, which helps to increase the robustness of our model against different disparities across views.
The evaluation is conducted on light fields from multiple sources, including real ones such as the New Light Field Image Dataset (EPFL)~\cite{Rerabek2016New,Mousnier2015Partial}, the synthetic ones such as HCI datasets~\cite{Honauer2016Dataset,Wanner2013Datasets}, the microscope datasets~\cite{Levoy2006Light,Lin2015Camera} that contain complex occlusions and translucency, and the camera gantry light fields such as the Gantry Archive~\cite{Wilburn2005High}. The experimental results demonstrate that the trained network can be generalized to various real-world scenes, synthetic scenes, and microscopy light fields. This shows that the geometric features are relatively representative of multiple situations.

\subsection{Model design}
In this section, we evaluate the model with 5 residual blocks in the spatio-angular restoration stage, and 3 residual blocks in the details refinement stage. Furthermore, to analyze the performance of our model, we vary the filter size and the local residual connections. We also analyze the effects of the multi-range training strategy.

\subsubsection{Angular filter size of 4D convolution}
\begin{figure}[t]
\centering
\includegraphics[width=0.9\columnwidth]{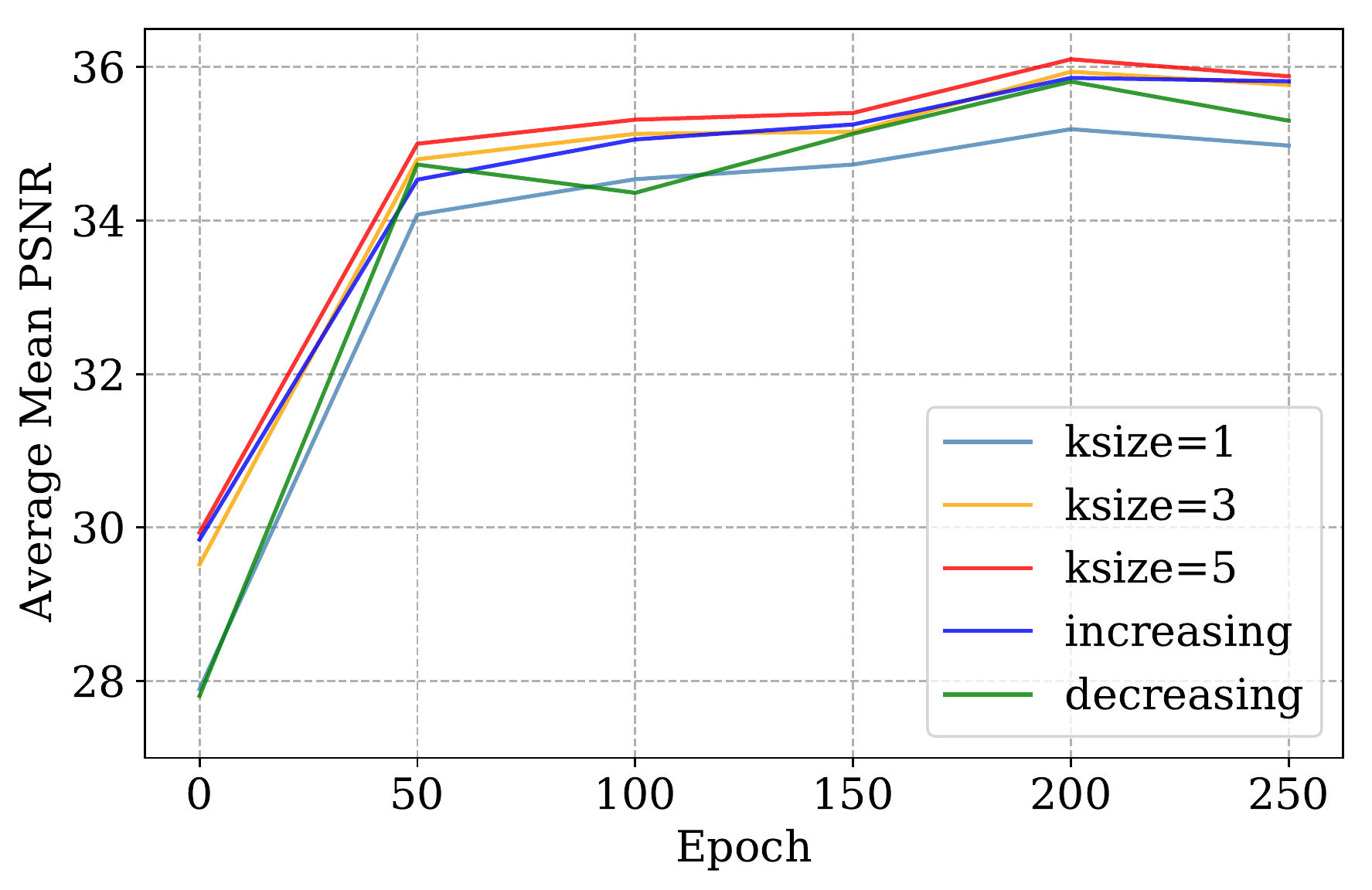}
\caption{Finding the angular kernel size. The curves are based on the average mean PSNR on a subset of the Stanford Archive scenes with spatial scaling factor $\times 2$ and angular scaling factor $\times 2$.}\label{fig:kernel_compare}
\end{figure}

To find the effective filter size to aggregate the angular information through the restoration network, we test five different settings of 4D convolution. We experiment with two types of architectures: (1) all convolution layers have the same kernel size ($1\times1$, $3\times3$, or $5\times5$); (2) the kernel size  increases ($1$--$1$--$3$--$3$--$5$) or decreases ($5$--$5$--$3$--$3$--$1$) across the layer. Note that for spatial super-resolution, varying the filter size does not have a significant impact on the performance, and certain filter size (e.g., $3\times3$) already can model the spatial content well~\cite{Simonyan2014Very}.
As is shown in Fig.~\ref{fig:kernel_compare}, our experiments show that the performance of networks with $5\times5$, $3\times3$ and increasing angular kernel size have competitive capacity to learn the angular correlations. 

\subsubsection{Varying residual connections}
\begin{figure}[t]
\centering
    \subfloat[][Sequential \\ skip connection]{
    \includegraphics[width=0.2\columnwidth,height=0.25\textheight]{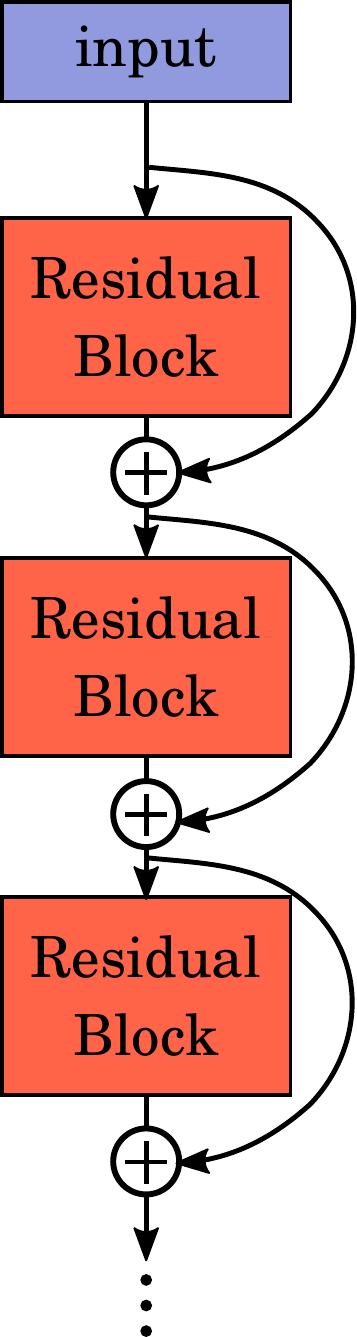}
    }
    \hspace{8mm}
    \subfloat[][Shared-source \\ skip connection]{
    \includegraphics[width=0.24\columnwidth,height=0.25\textheight]{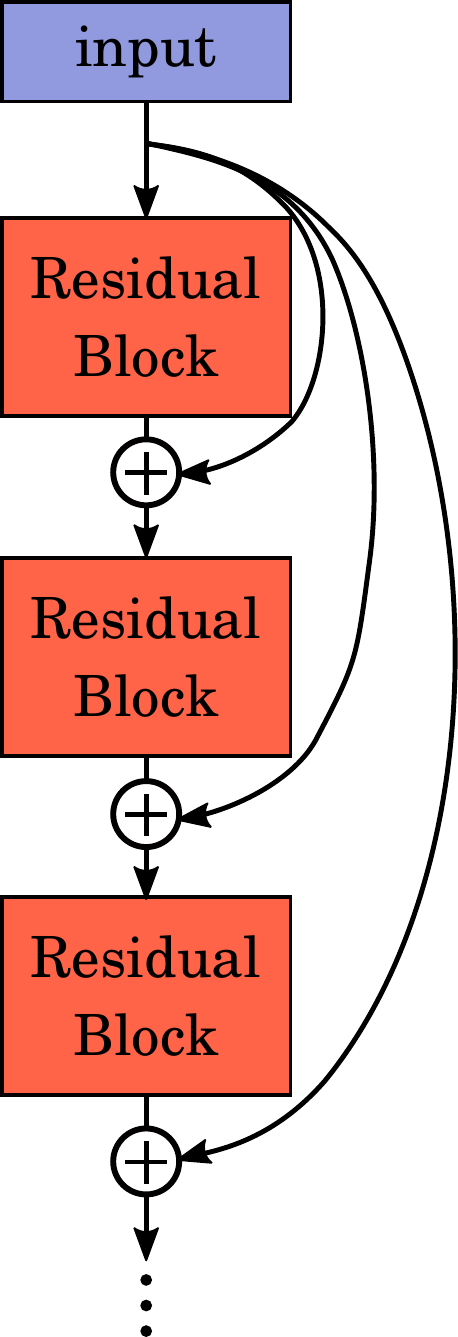}
    }
    \hspace{5mm}
    \subfloat[][Dense skip \\ connection]{
    \includegraphics[width=0.24\columnwidth,height=0.25\textheight]{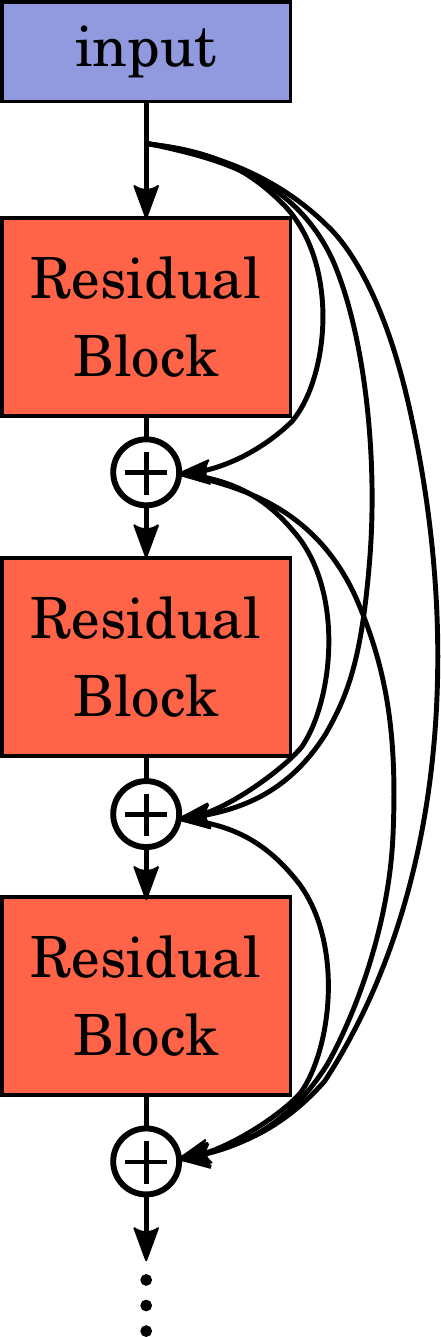}
    }
    \caption{\textbf{Local residual connection.} We explore three different ways of skip connection in the residual modules for training the proposed models.}\label{fig:connection}
\end{figure}

\begin{figure}[t]
\centering
\includegraphics[width=0.9\columnwidth]{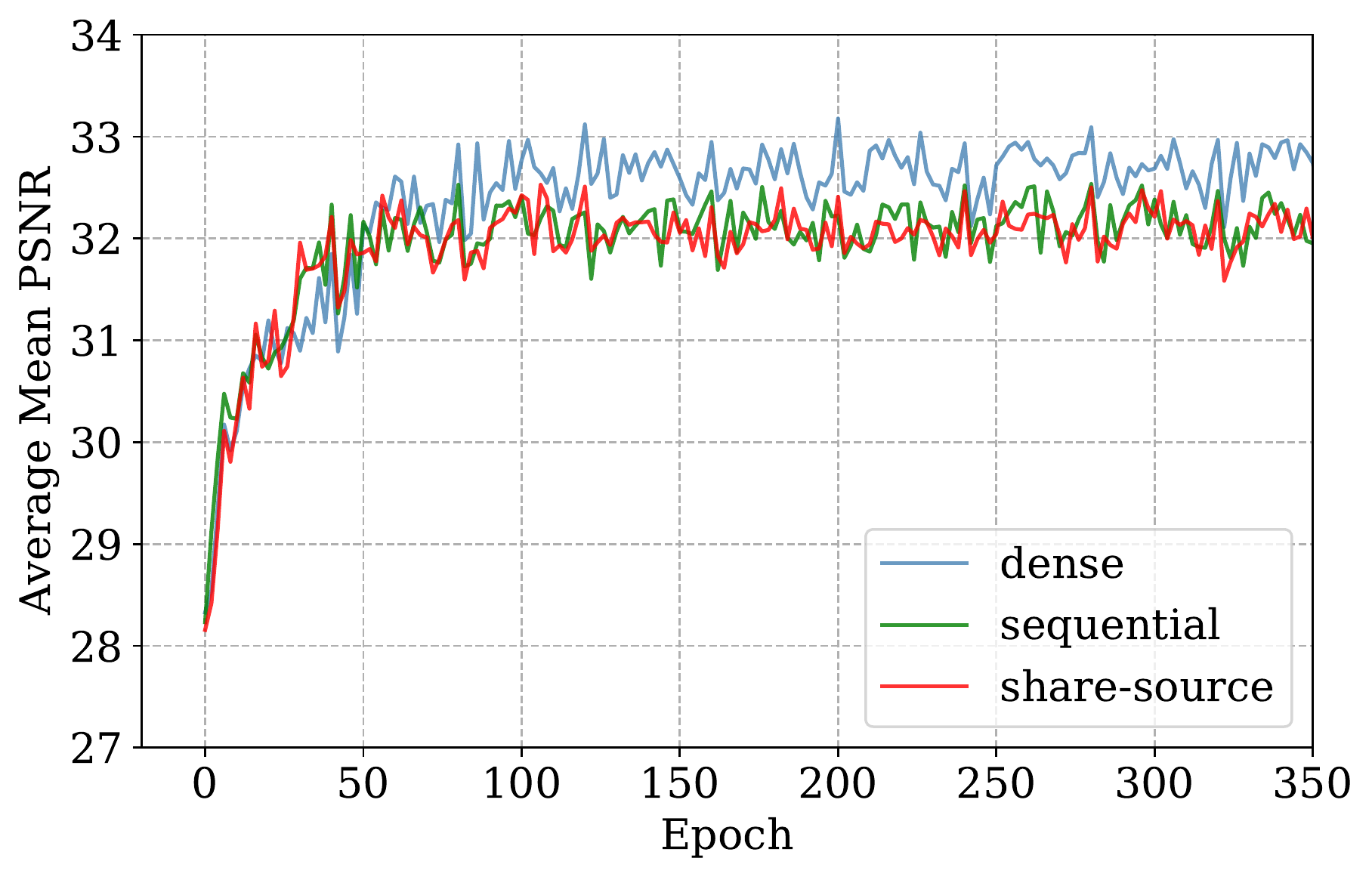}
\caption{Convergence analysis on different types of connections. The curves for each connection are based on the Average Mean PSNR on the validation set.}\label{fig:connection_convergence}
\end{figure}

We examine three different methods of local residual learning in our model to evaluate the effects of hierarchical spatial-angular features from the original LR light fields.
\begin{enumerate}
\item \textbf{Sequential skip connection}: This is the classic connection style used in ResNet. We adopt and extend the method to fit our high-dimension convolution layer.
\item \textbf{Shared-source skip connection}: All residual blocks are connected to the source features.
\item \textbf{Dense skip connection}: A dense-style connection motivated by DenseNet~\cite{Huang2017Densely}, which makes full use of hierarchical spatio-angular features.
\end{enumerate}
We illustrate the three types of connection in Fig.~\ref{fig:connection}, and Fig.~\ref{fig:connection_convergence} shows the convergence curves of each type of connection. The dense-skip connection ensures that the model converges to a better point.

\subsubsection{Reconstruction strategy}
\begin{figure}[!ht]
    \centering
    \includegraphics[width=1.\columnwidth]{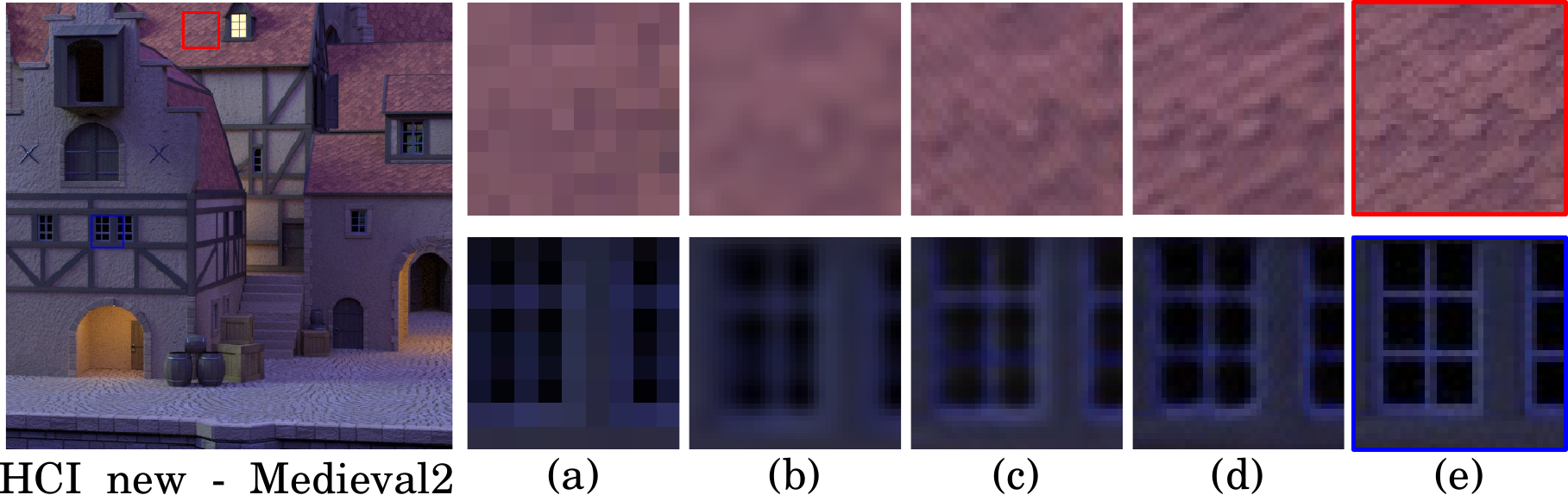}
    \caption{\textbf{Contribution of different loss.} (a) The input LR LF image (b) Bicubic (c) Model R5D3 trained on angular loss (d) Model R5D3 trained on the combination of spatial and angular loss (e) Ground-truth}
    \label{fig:ablation_perceptual}
\end{figure}

We compare different variants of the proposed model with different loss items and multi-range training strategy. Our model is composed of two parts, namely spatio-angular information reconstruction represented by ``R'', and spatial details refinement represented by ``D''. Table~\ref{table:ablation_components} compares the effects of different components, together with the learning strategy and loss for $2\times$ spatial and $2\times$ angular SR task. The number behind R and D represents the number of HD residual blocks having been involved in the corresponding subnetwork. ``MR'' stands for multi-range training, and ``Refinement'' denotes whether the model contains the details refinement part. The quantitative results show performance improvement, which validates the effectiveness of multi-range training strategy and the defined loss. 
Fig.~\ref{fig:ablation_perceptual} examines the contribution of using the spatial loss function on the high-frequency details reconstruction. As is discussed in Section~\ref{sec:loss_function}, this loss function helps to promote the restoration of high-frequency details (e.g.\ the roof tile texture and window frame).

\begin{table}[!ht]
\setlength{\tabcolsep}{0.5em}
\def\arraystretch{1.2}
\centering
\caption{Ablation study of different components in the proposed model. We compare the performs of several variants of the model on the HCI new test dataset and occlusion scenes, and report the PSNR results.}\label{table:ablation_components}
\begin{tabular}{cccc|cc}
\hline
Model & MR           & Refinement   & Loss            & HCI new (test)  & Occlusion 10    \\ \hline
R8    & $\times$     & $\times$     & $\ell_A$        & $31.35$ & $32.70$ \\
D8    & $\times$     & $\checkmark$ & $\ell_S$        & $31.34$ & $32.65$ \\
R5D3  & $\times$     & $\checkmark$ & $\ell_A+\ell_S$ & $31.34$ & $32.76$ \\
R8    & $\checkmark$ & $\times$     & $\ell_A$        & $31.64$ & $32.77$ \\
D8    & $\checkmark$ & $\checkmark$ & $\ell_S$        & $31.65$ & $\mathbf{33.30}$ \\
R5D3  & $\checkmark$ & $\checkmark$ & $\ell_A+\ell_S$ & $\mathbf{31.74}$ & $33.29$ \\ \hline
\end{tabular}
\end{table}
\vspace{-1em}

\section{Results}
\label{sec:results}
\begin{table*}[t]
\caption{Quantitative evaluation of state-of-the-art LFSR algorithms. We report the average PSNR and SSIM for Spatial $2\times$, $3\times$, $4\times$ and Angular $2\times$, $3\times$. \fst{Red} and \blue{blue} indicate the best and the second best performance, respectively.}
\label{table:overall_quatitive_evaluations}
\setlength{\tabcolsep}{2.2pt}
\centering
\begin{tabular}{|r|c|ccccccc|ccccccc|}
\hline
\multicolumn{1}{|c|}{\multirow{2}{*}{Algorithm}} & \multirow{2}{*}{Scale}        & \multicolumn{7}{c|}{PSNR (dB)}                                                                                                                                                                                                                                                                               & \multicolumn{7}{c|}{SSIM}                                                                                                                                                                                                                                                                               \\ \cline{3-16} 
\multicolumn{1}{|c|}{}                           &                               & \begin{tabular}[c]{@{}c@{}}Occlusions\\ (20)\end{tabular} & \begin{tabular}[c]{@{}c@{}}Reflective\\ (20)\end{tabular} & \begin{tabular}[c]{@{}c@{}}HCI\\ old\end{tabular} & \begin{tabular}[c]{@{}c@{}}HCI\\ new\end{tabular} & Micro. & Stanford & \begin{tabular}[c]{@{}c@{}}EPFL\\ (21)\end{tabular} & \begin{tabular}[c]{@{}c@{}}Occlusions\\ (20)\end{tabular} & \begin{tabular}[c]{@{}c@{}}Reflective\\ (20)\end{tabular} & \begin{tabular}[c]{@{}c@{}}HCI\\ old\end{tabular} & \begin{tabular}[c]{@{}c@{}}HCI\\ new\end{tabular} & Micro. & Stanford & \begin{tabular}[c]{@{}c@{}}EPFL\\ (21)\end{tabular} \\ \hline \hline
Bicubic                                          & \multirow{11}{*}{S$\times 2$} &   28.52   &   31.19   &   26.97   &   29.69   &   30.13   &   31.75   &   28.66      &   0.808   &   0.863   &   0.769   &   0.806   &   0.797   &   0.919   &   0.849   \\
Yoon et al.~\cite{Yoon2017Light}                 &                               &   28.86   &   31.42   &   28.41   &   31.24   &   30.99   &   32.71   &   29.69      &   0.834   &   0.885   &   0.826   &   0.851   &   0.798   &   0.937   &   0.864   \\
BM PCA+RR~\cite{Farrugia2017Super}               &                               &   30.45   &   33.07   &   31.20   &   32.62   &   28.93   &   32.92   &   32.68      &   0.878   &   0.896   &   0.892   &   0.879   &   0.731   &   0.921   &   0.906   \\
LFNet~\cite{Wang2018Lfnet}                       &                               &   30.37   &   33.85   &   29.56   &   32.81   &   30.11   &   32.21   &   32.66      &   0.881   &   0.912   &   0.884   &   0.898   &   0.813   &   0.924   &   0.892   \\
VDSR~\cite{Kim2016Accurate}                      &                               &   29.84   &   32.32   &   29.32   &   32.11   &   31.28   &   32.29   &   30.46      &   0.865   &   0.898   &   0.821   &   0.886   &   0.839   &   0.940   &   0.892   \\
MSLapSRN~\cite{Lai2017Fast}                      &                               &   30.85   &   32.43   &   29.51   &   33.13   &   31.63   &   34.07   &   32.27      &   0.879   &   0.909   &   0.852   &   0.894   &\snd{0.867}&   0.945   &   0.901   \\
RDN~\cite{Zhang2018Residual}                     &                               &   31.46   &   33.86   &   31.13   &   33.25   &   31.91   &   33.22   &   32.41      &   0.893   &   0.916   &   0.894   &   0.893   &   0.858   &   0.893   &   0.912   \\
ESPCN~\cite{Shi2016Real}                         &                               &   32.72   &   35.38   &   31.18   &   33.42   &   32.19   &   36.61   &   32.41      &   0.911   &   0.936   &   0.900   &   0.896   &   0.843   &   0.963   &   0.926   \\
Jo et al.~\cite{Jo2018Deep}                      &                               &   32.29   &   34.54   &   30.92   &   32.93   &   32.59   &   34.75   &   31.83      &   0.902   &   0.920   &   0.871   &   0.860   &   0.866   &   0.946   &   0.909   \\
\textbf{HDDRNet}                                 &                               &\snd{34.69}&\snd{36.34}&\snd{32.64}&\snd{34.20}&\snd{32.64}&\snd{37.49}&\snd{35.30}   &\snd{0.934}&\snd{0.949}&\snd{0.932}&\fst{0.916}&\fst{0.872}&\snd{0.967}&\snd{0.945}\\
\textbf{M-HDDRNet}                               &                               &\fst{34.83}&\fst{37.06}&\fst{33.12}&\fst{34.64}&\fst{33.00}&\fst{38.30}&\fst{35.97}   &\fst{0.935}&\fst{0.950}&\fst{0.933}&\snd{0.915}&   0.866   &\fst{0.969}&\fst{0.947}\\ \hline \hline

Bicubic                                          & \multirow{11}{*}{S$\times 3$} &   26.95   &   29.50   &   25.39   &   28.94   &   29.04   &   29.42   &   27.31      &   0.746   &   0.819   &   0.703   &   0.776   &   0.760   &   0.895   &   0.812   \\
Yoon et al.~\cite{Yoon2017Light}                 &                               &   27.25   &   29.78   &   25.79   &   29.57   &   29.06   &   29.78   &   27.51      &   0.758   &   0.826   &   0.733   &   0.803   &   0.724   &   0.912   &   0.827   \\
BM PCA+RR~\cite{Farrugia2017Super}               &                               &   27.96   &   30.05   &   26.78   &   30.24   &   29.00   &   29.96   &   29.71      &   0.817   &   0.835   &   0.766   &   0.834   &   0.732   &   0.875   &   0.850   \\
LFNet~\cite{Wang2018Lfnet}                       &                               &   28.01   &   30.34   &   26.76   &   29.81   &   29.34   &   29.98   &   29.69      &   0.816   &   0.847   &   0.764   &   0.822   &   0.741   &   0.864   &   0.848   \\
VDSR~\cite{Kim2016Accurate}                      &                               &   27.94   &   29.77   &   26.28   &   29.44   &   29.38   &   29.67   &   27.62      &   0.809   &   0.841   &   0.721   &   0.803   &   0.739   &   0.873   &   0.855   \\
MSLapSRN~\cite{Lai2017Fast}                      &                               &   29.22   &   32.03   &   27.80   &   30.94   &   30.21   &   33.61   &   30.00      &   0.818   &   0.875   &   0.789   &   0.828   &   0.758   &   0.935   &   0.867   \\
RDN~\cite{Zhang2018Residual}                     &                               &   29.14   &   30.82   &   26.89   &   29.54   &   29.68   &   32.92   &   29.65      &   0.796   &   0.846   &   0.755   &   0.788   &   0.761   &   0.891   &   0.834   \\
ESPCN~\cite{Shi2016Real}                         &                               &   28.90   &   31.47   &   27.46   &   29.96   &   30.10   &   33.36   &   30.30      &   0.809   &   0.866   &   0.786   &   0.819   &   0.780   &\snd{0.938}&   0.855   \\
Jo et al.~\cite{Jo2018Deep}                      &                               &   30.62   &   33.19   &   29.05   &   31.76   &   30.11   &   33.25   &   30.86      &   0.859   &   0.896   &   0.822   &   0.852   &\snd{0.793}&   0.934   &   0.891   \\
\textbf{HDDRNet}                                 &                               &\fst{31.18}&\snd{33.34}&\fst{29.53}&\snd{31.97}&\snd{30.59}&\snd{33.74}&\snd{32.68}   &\snd{0.872}&\fst{0.902}&\snd{0.848}&\snd{0.865}&   0.786   &\snd{0.938}&\fst{0.904}\\
\textbf{M-HDDRNet}                               &                               &\snd{31.08}&\fst{33.37}&\snd{29.41}&\fst{32.11}&\fst{30.93}&\fst{34.03}&\fst{32.73}   &\fst{0.875}&\fst{0.902}&\fst{0.855}&\fst{0.869}&\fst{0.807}&\fst{0.940}&\fst{0.904}\\ \hline \hline

Bicubic                                          & \multirow{11}{*}{S$\times 4$} &   24.98   &   27.54   &   23.95   &   25.92   &   27.46   &   26.99   &   25.94      &   0.663   &   0.771   &   0.630   &   0.688   &   0.705   &   0.842   &   0.767   \\
Yoon et al.~\cite{Yoon2017Light}                 &                               &   25.04   &   28.14   &   25.65   &   28.28   &   28.02   &   29.25   &   26.97      &   0.686   &   0.798   &   0.688   &   0.768   &   0.710   &   0.860   &   0.792   \\
BM PCA+RR~\cite{Farrugia2017Super}               &                               &   26.28   &   28.73   &   25.85   &   28.90   &   27.32   &   29.91   &   27.51      &   0.710   &   0.796   &   0.703   &   0.772   &   0.675   &   0.865   &   0.785   \\
LFNet~\cite{Wang2018Lfnet}                       &                               &   25.94   &   28.81   &   25.40   &   29.36   &   28.21   &   28.67   &   26.10      &   0.709   &   0.808   &   0.706   &   0.762   &   0.718   &   0.835   &   0.775   \\
VDSR~\cite{Kim2016Accurate}                      &                               &   25.00   &   27.72   &   25.21   &   29.05   &   28.23   &   29.38   &   25.99      &   0.671   &   0.780   &   0.672   &   0.765   &   0.722   &   0.863   &   0.772   \\
MSLapSRN~\cite{Lai2017Fast}                      &                               &   27.41   &   30.28   &   26.27   &   29.55   &   29.13   &   31.70   &   28.78      &   0.755   &   0.835   &   0.723   &   0.782   &   0.715   &\fst{0.907}&   0.821   \\
RDN~\cite{Zhang2018Residual}                     &                               &   26.97   &   29.64   &   26.66   &   29.63   &   28.80   &\snd{31.81}&   28.58      &   0.724   &   0.817   &   0.730   &   0.792   &   0.742   &   0.895   &   0.804   \\
ESPCN~\cite{Shi2016Real}                         &                               &   27.14   &   29.84   &   26.07   &   29.06   &\snd{29.22}&   30.58   &   27.95      &   0.745   &   0.826   &   0.717   &   0.771   &\snd{0.744}&   0.890   &   0.812   \\
Jo et al.~\cite{Jo2018Deep}                      &                               &   27.59   &   30.32   &   26.72   &   29.75   &\fst{29.63}&   30.53   &   28.52      &   0.765   &   0.838   &   0.722   &   0.787   &\fst{0.769}&   0.891   &   0.830   \\
\textbf{HDDRNet}                                 &                               &\snd{28.40}&\snd{30.57}&\snd{27.66}&\snd{29.83}&   28.60   &   30.91   &\snd{29.97}   &\snd{0.790}&\snd{0.846}&\snd{0.789}&\snd{0.814}&   0.693   &   0.891   &\snd{0.846}\\
\textbf{M-HDDRNet}                               &                               &\fst{28.70}&\fst{30.78}&\fst{27.97}&\fst{30.46}&   28.98   &\fst{31.94}&\fst{30.61}   &\fst{0.805}&\fst{0.853}&\fst{0.801}&\fst{0.827}&   0.708   &\fst{0.907}&\fst{0.862}\\ \hline \hline

Yoon et al.~\cite{Yoon2017Light}                 & \multirow{5}{*}{A$\times 2$}  &   34.55   &   35.30   &   30.65   &   33.54   &   32.43   &   33.75   &   35.21      &   0.910   &   0.939   &   0.779   &   0.892   &   0.905   &   0.849   &   0.939   \\
Kalantari et al.~\cite{Kalantari2016Learning}    &                               &   36.50   &   38.73   &   32.95   &   36.96   &   33.87   &   33.37   &   38.70      &   0.943   &   0.969   &   0.915   &\fst{0.933}&   0.910   &\fst{0.936}&   0.972   \\
Wu et al.~\cite{Wu2018Light}                     &                               &   37.76   &   40.36   &   33.62   &   36.24   &   35.85   &   34.96   &   39.37      &   0.952   &   0.970   &\fst{0.920}&   0.924   &\fst{0.944}&   0.901   &\fst{0.973}   \\
\textbf{HDDRNet}                                 &                               &\snd{38.27}&\snd{41.22}&\snd{35.56}&\snd{37.72}&\snd{37.21}&\snd{35.05}&\snd{40.02}   &\fst{0.953}&\snd{0.972}&   0.918   &\snd{0.925}&   0.932   &\snd{0.908}&\fst{0.973}\\
\textbf{M-HDDRNet}                               &                               &\fst{38.45}&\fst{41.38}&\fst{35.77}&\fst{37.90}&\fst{37.39}&\fst{35.27}&\fst{40.22}   &\fst{0.953}&\fst{0.973}&\snd{0.919}&   0.924   &\snd{0.934}&   0.904   &\fst{0.973}\\ \hline \hline

Kalantari et al.~\cite{Kalantari2016Learning}    & \multirow{4}{*}{A$\times 3$}  &   34.70   &   37.24   &   32.59   &   35.53   &   30.71   &   28.84   &   35.19      &   0.927   &   0.958   &\snd{0.906}&\snd{0.916}&   0.826   &   0.850   &   0.959   \\
Wu et al.~\cite{Wu2018Light}                     &                               &   35.64   &   40.03   &   33.38   &   35.64   &   31.08   &   30.21   &   37.05      &\snd{0.928}&   0.963   &   0.905   &\fst{0.918}&\snd{0.845}&   0.852   &   0.960   \\
\textbf{HDDRNet}                                 &                               &\snd{35.96}&\fst{40.16}&\snd{33.75}&\snd{35.79}&\snd{31.34}&\snd{30.24}&\snd{38.28}   &\snd{0.928}&\fst{0.964}&   0.905   &   0.904   &\fst{0.847}&\snd{0.857}&\snd{0.961}\\
\textbf{M-HDDRNet}                               &                               &\fst{36.05}&\snd{40.14}&\fst{34.23}&\fst{36.45}&\fst{31.38}&\fst{30.61}&\fst{38.41}   &\fst{0.929}&\fst{0.964}&\fst{0.913}&\snd{0.916}&   0.842   &\fst{0.861}&\fst{0.962}\\ \hline
\end{tabular}
\end{table*}

\subsection{Overall Comparisons}
For comparisons in terms of spatial resolution, we choose 8 state-of-the-art SR algorithms, including 3 LFSR methods (Yoon et al.~\cite{Yoon2017Light}, BM PCA+RR~\cite{Farrugia2017Super}, LFNet~\cite{Wang2018Lfnet}), 3 well-known single-image SR methods (VDSR~\cite{Kim2016Accurate}, MSLapSRN~\cite{Lai2017Fast}, RDN~\cite{Zhang2018Residual}) and 2 video SR (ESPCN~\cite{Shi2016Real} and Jo et al.~\cite{Jo2018Deep}). We examine different methods on 7 public light field datasets on real-world, synthetic and microscope data. The results are evaluated with the widely used image quality metrics: peak signal-to-noise ratio (PSNR) and structural similarity (SSIM), comparing the performance on $2\times$, $3\times$ and $4\times$ SR. For each comparison, we use the public source code and fine-turning of the model to fit the classic downsampling method described in Eq.~\ref{equ:downsampling}. (\ul{for details, please refer to supplementary materials}) The quantitative results are shown in Table~\ref{table:overall_quatitive_evaluations}. Our M-HDDRNet performs favorably against existing 2D and 3D (video) SR methods. One major limitation of these methods is that they do not fully exploit the light field structure, where each sub-aperture image is restored independently (2D) or with 1D correlations. However, the sub-aperture images in light field are correlated in 2D, and our M-HDDRNet is able to fully exploit such complex angular correlations to reconstruct high-quality scenes.

For angular SR, we compare with several recent learning-based methods on different tasks. In Table~\ref{table:overall_quatitive_evaluations}, we evaluate the performance against methods proposed by Kalantari et al.~\cite{Kalantari2016Learning} and Wu et al.~\cite{Wu2018Light} on real-world, synthetic and microscope datasets. Both PSNR and SSIM are presented in Table~\ref{table:overall_quatitive_evaluations}, where ``A$\times2$'' refers to enhance angular resolution from $5 \times 5$ to $9 \times 9$, and ``A$\times3$'' means enhance angular resolution from $3 \times 3$ to $9 \times 9$.

\begin{figure*}
\centering
\includegraphics[width=0.9\textwidth]{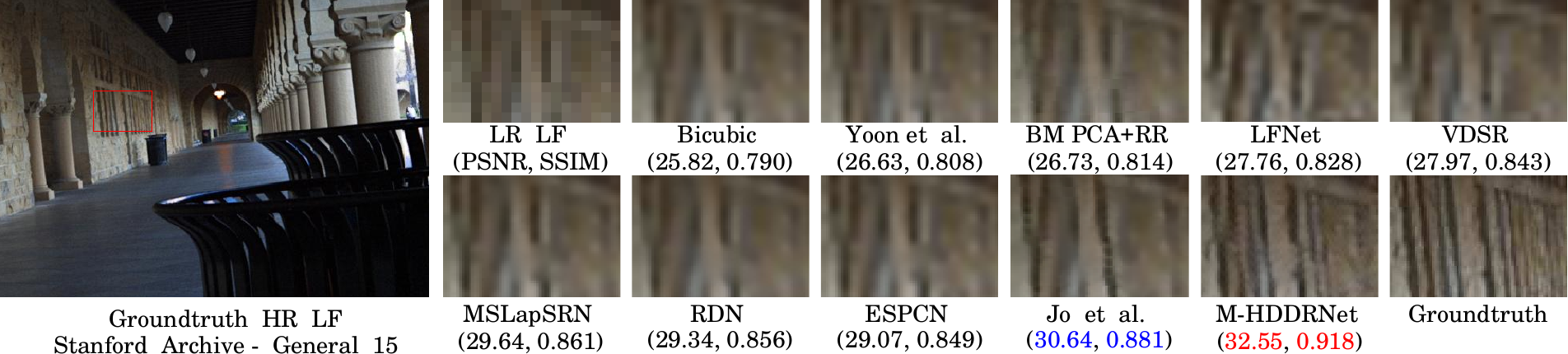}\vspace{1mm}
\includegraphics[width=0.9\textwidth]{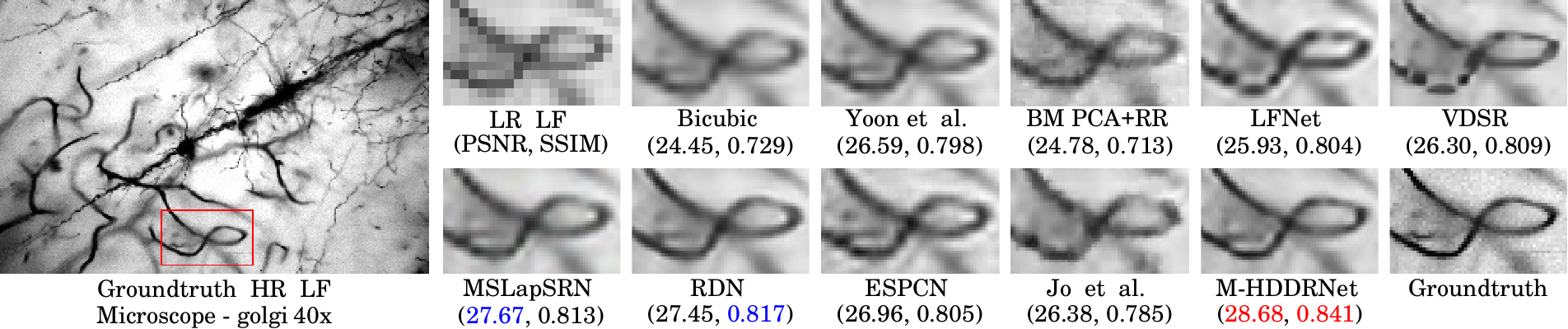}\vspace{1mm}
\includegraphics[width=0.9\textwidth]{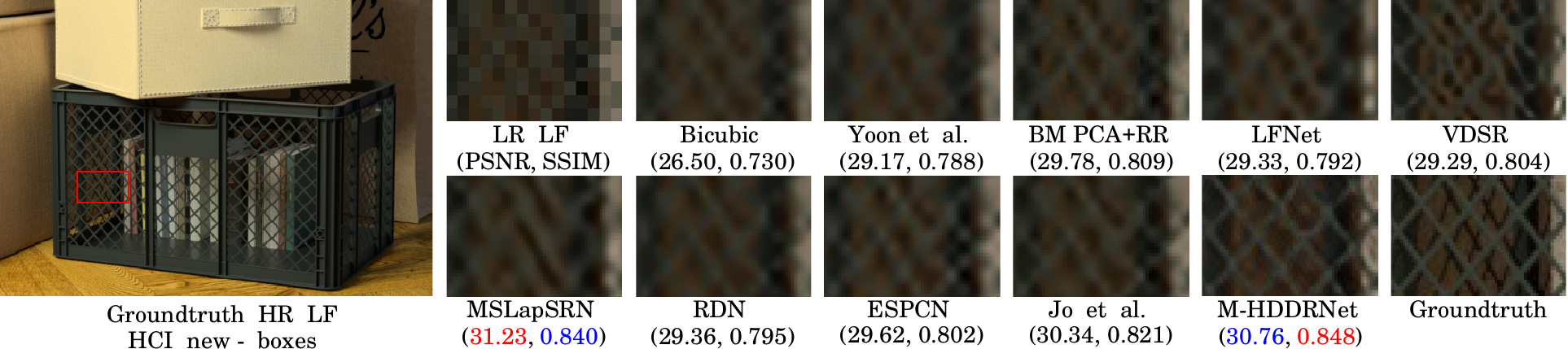}
\caption{Visual comparison for $4\times$ SR on the Stanford Archive (real-world), $3\times$ SR on Microscope, and $4\times$ SR on HCI new (synthetic) dataset.}\label{fig:result_general15}
\end{figure*}

\subsubsection{Comparison with 2D SR methods}
\begin{figure}[!t]
\centering
\includegraphics[width=1.0\columnwidth]{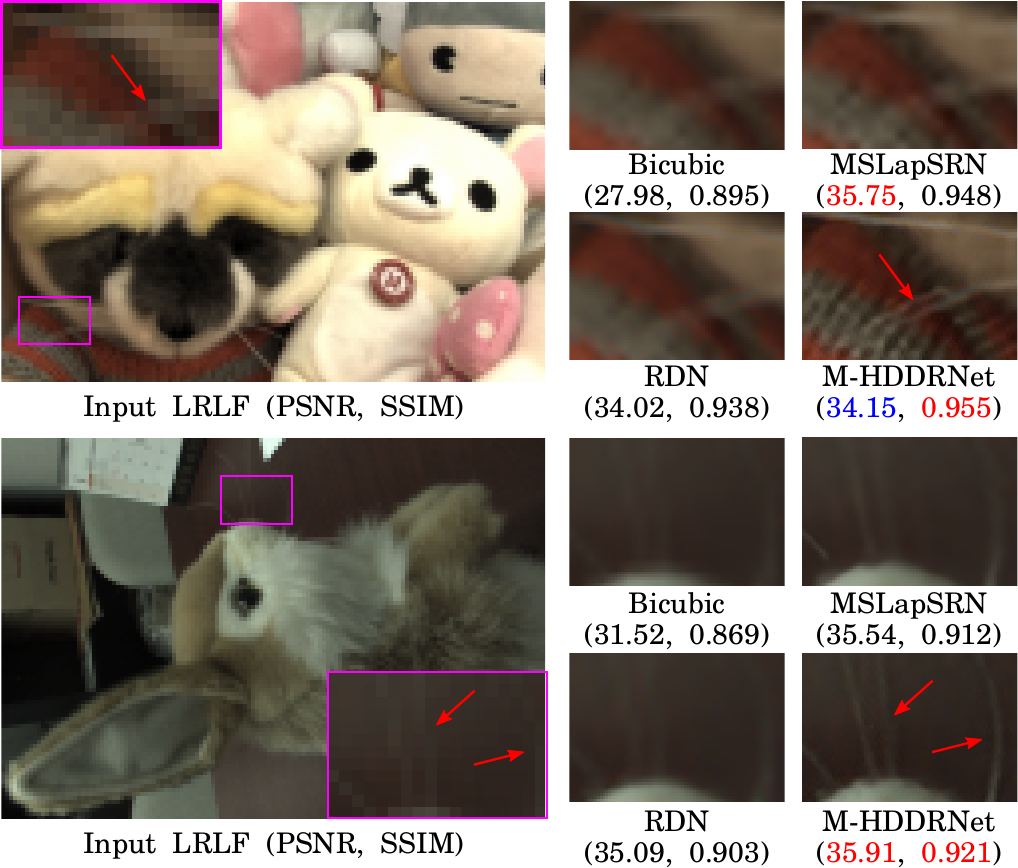}
\caption{\textbf{Comparison with 2D image SR methods on real-world LF scene for $4\times$ SR.} The LRLF loses the detailed texture after downsampling, and our model super-resolves the ``whisker'' accurately while MSLapSRN~\cite{Lai2017Fast} and RDN~\cite{Zhang2018Residual} failed to reconstruct such fine texture information.}\label{fig:result_subpixel_LF067_LF006}
\end{figure}

To illustrate the benefits of using high-dimensional convolution, we compare the visual performance on the scenes containing fine structures with two state-of-the-art 2D SR methods. As is shown in Fig.~\ref{fig:result_subpixel_LF067_LF006}, our proposed model successfully restores the fine texture (the ``whisker'') that is almost lost in the LR input scenes. The results generated from 2D SR methods~\cite{Lai2017Fast} and~\cite{Zhang2018Residual} are blurry, especially on the ``whisker'' regions, even if they have competitive PSNR and SSIM with ours.

\subsubsection{Comparison with 3D SR methods}
\begin{figure}[!t]
\centering
\includegraphics[width=1.0\columnwidth]{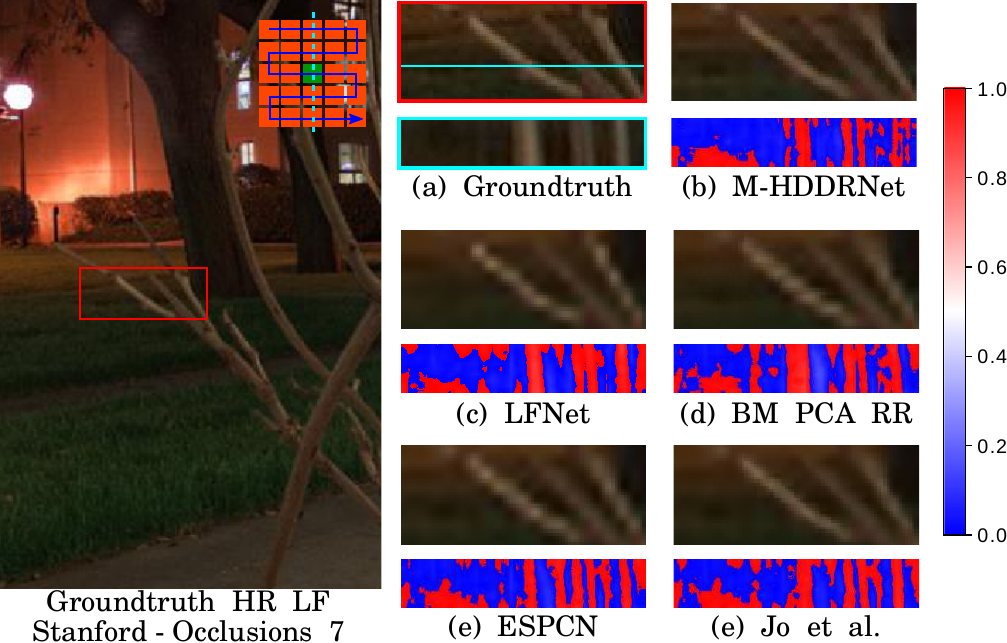}
\caption{\textbf{Comparison with 3D image SR methods on real-world LF scene for $4\times$ SR.} We present both the spatial SR results (center view) and the error EPI of the focused region.}\label{fig:result_3d_compaison}
\end{figure}

3D SR methods on LFSR treat the light field as a sequence of images, and therefore they all lose 1D angular correlation. In our experiment, we rearrange the sub-aperture images of a set of light field as an image sequence, and compare the 
results of our model with other state-of-the-art 3D SR methods in Fig.~\ref{fig:result_3d_compaison}. M-HDDRNet gives more realistic spatial results while preserving good angular correlations in 2D.

\subsection{\texorpdfstring{$2\times2$ to $8\times8$}{Lg} view synthesis comparison}
\begin{figure*}[!ht]
\centering
\includegraphics[width=0.8\textwidth]{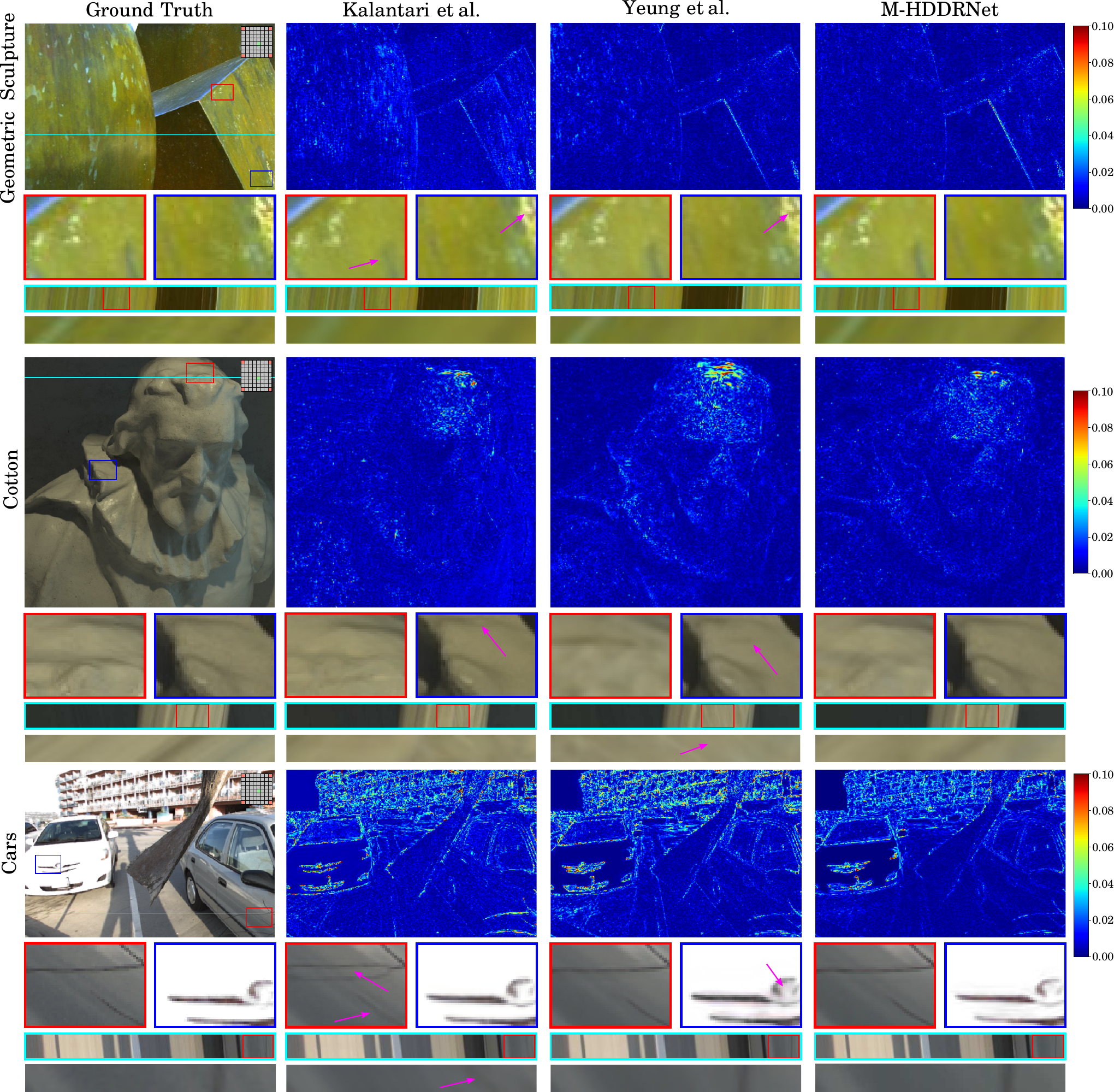}
\caption{Visual comparison of our model with Kalantari et al.~\cite{Kalantari2016Learning} and Yeung et al.~\cite{Yeung2018Fast} for $2\times2$ -- $8\times8$ angular SR task. The first column presents the ground truth LF, and the other columns show the residual results between the reconstruction LF and ground truth LF on the $(5,5)$ synthesized sub-aperture image. We zoom in some regions for better comparison.}\label{fig:result_angular}
\end{figure*}

In this section, we carried out comparison with two state-of-the-art view synthesis methods, namely Kalantari et al.~\cite{Kalantari2016Learning} and Yeung et al.~\cite{Yeung2018Fast}. The method by Wu et al.~\cite{Wu2018Light} cannot be compared since their method requires 3 views in each angular dimension to provide enough information for interpolation step. Table~\ref{table:2x2_8x8_angular} shows the average performance on several public LF datasets and our model obtains higher PSNR value than the other two methods. Fig.~\ref{fig:result_angular} further visually demonstrates that our model is able to obtain better reconstruction quality. Kalantari et al.~\cite{Kalantari2016Learning} tends to produce artifacts near the boundaries, especially in the region with complex occlusions. Our model reconstructs the LF preserved better geometric structure by fully using the correlations among sub-aperture images.

\begin{table}[!ht]
\caption{Quantitative evaluation of state-of-the-art view synthesis algorithms. We report the average PSNR under the task $2\times2-8\times8$.}
\label{table:2x2_8x8_angular}
\setlength{\tabcolsep}{2.5pt}
\centering
\begin{tabular}{r|ccccc}
\hline
Algorithm                                      & \begin{tabular}[c]{@{}c@{}}Occlusions\\ (20)\end{tabular} & \begin{tabular}[c]{@{}c@{}}Reflective\\ (20)\end{tabular} & \begin{tabular}[c]{@{}c@{}}EPFL\\ (21)\end{tabular} & Micro. & \begin{tabular}[c]{@{}c@{}}HCI\\ new\end{tabular} \\ \hline
Kalantari et al.~\cite{Kalantari2016Learning}  &  32.68           &  35.98           &  33.60           &  22.14           &  33.19  \\
Yeung et al.(16L)~\cite{Yeung2018Fast}         &  33.19           &  36.82           &  35.09           &  24.11           &  33.39  \\
\textbf{M-HDDRNet}                             &  \textbf{33.24}  &  \textbf{36.97}  &  \textbf{35.34}  &  \textbf{24.13}  &  \textbf{34.04}  \\ \hline
\end{tabular}
\end{table}

\begin{figure*}[!ht]
\centering
\includegraphics[]{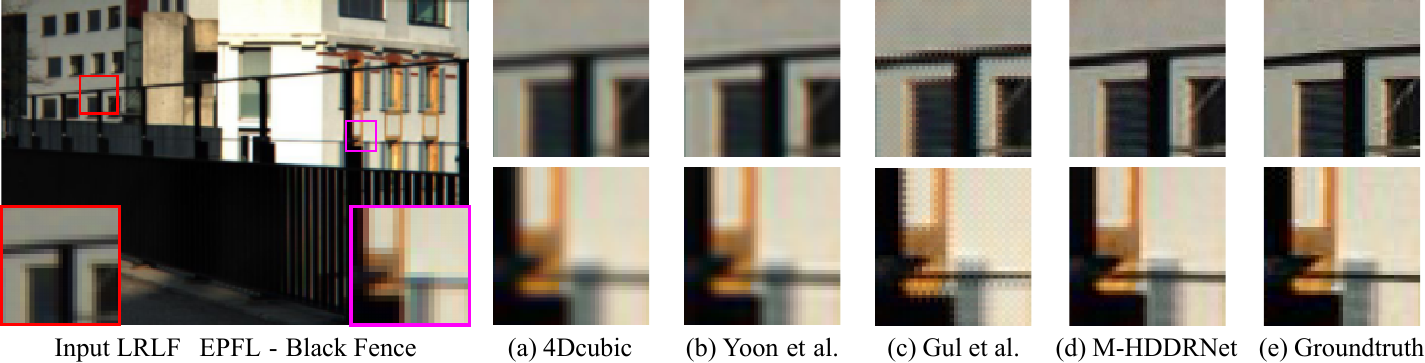}
\caption{Comparison with spatio-angular SR methods on real-world LF scene for $2\times$ on spatial and $2\times$ on angular resolution enhancement.}\label{fig:result_2s2a}
\end{figure*}

\subsection{Comparison with spatio-angular SR methods}
One major benefits of our proposed model is its capability to enhance both spatial and angular resolution simultaneously. We compare with two existing methods~\cite{Yoon2017Light,Gul2018Spatial} for super-resolution on both spatial and angular dimensions in Fig.~\ref{fig:result_2s2a}. Yoon et al.~\cite{Yoon2017Light} applied the SRCNN to recover spatial details leading to smooth results; Gul et al.~\cite{Gul2018Spatial} provided a pixel-level reconstruction strategy, recovering both spatial and angular information separately. However, such pixel-level approach easily results in lattice artifacts in bright regions of the scene, such as what is shown in the ``wall region'' of Fig.~\ref{fig:result_2s2a}\red{(c)}.

\subsection{Computational analysis}
\label{sec:computations}
\begin{figure}[!ht]
    \centering
    \includegraphics[width=0.95\columnwidth]{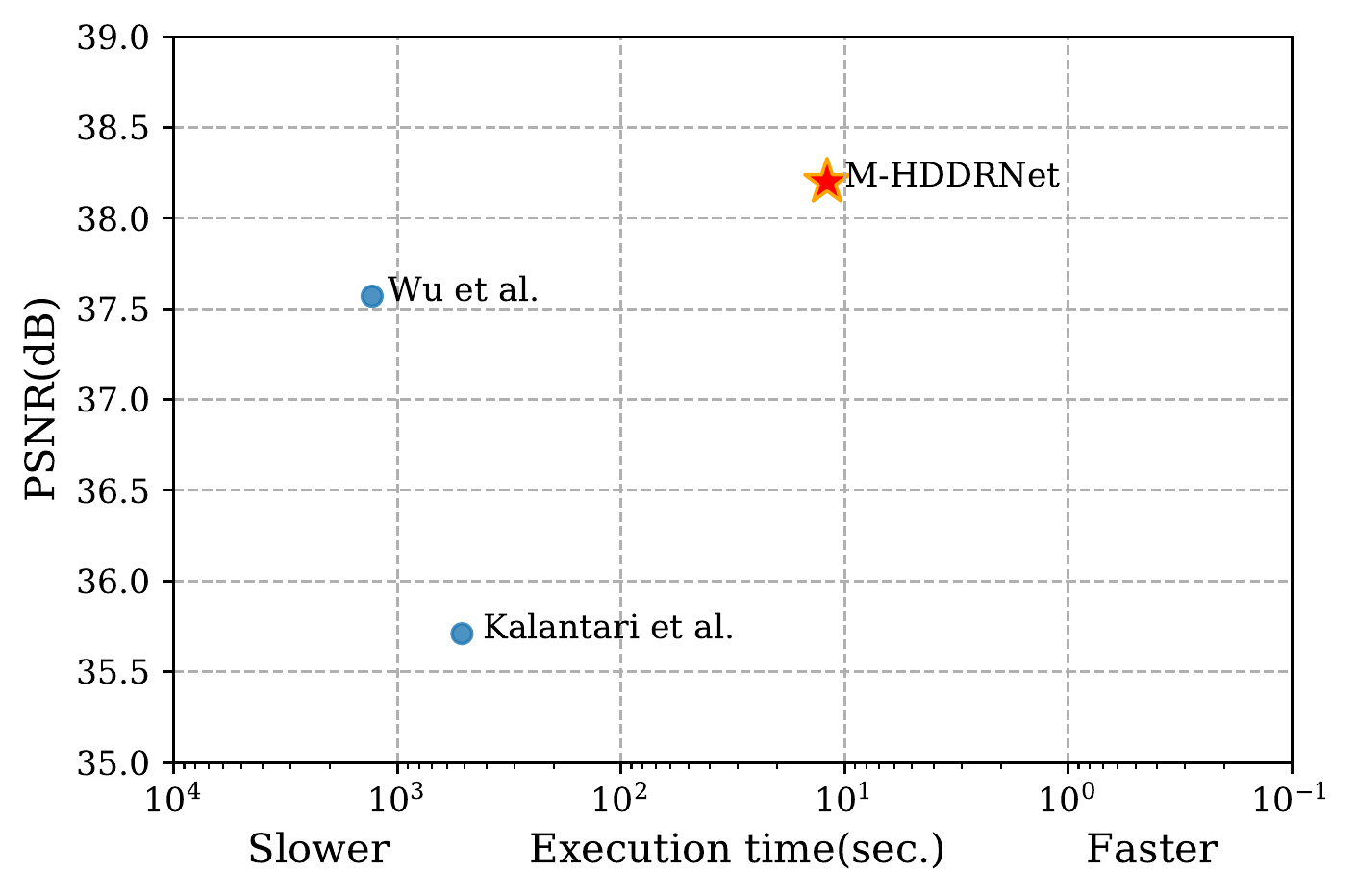}
    \caption{The runtime comparison of multiple schemes for $3\times3 \rightarrow 9\times9$ task. The execution time (sec.) of different frameworks are calculated on the same machine with 2.4GHz Intel i7 CPU and NVIDIA Titan X GPU (12G Memory) and the PSNR values are calculated over the average of 60 real-world test scenes.}
    \label{fig:computations}
\end{figure}
One potential drawback of the proposed 4D framework is its additional computational requirement for the demanding 4D convolution operations. Theoretically, when compare with a conventional 2D convolution layer, the proposed 4D convolution layer incur $a_1 \times a_2$ times additional compute operations to result in the same output size. However, as the 4D convolution allows our model to reconstruct the entire LF directly without additional auxiliary steps, the end-to-end compute time of our proposed framework remains competitive. To evaluate its end-to-end performance, we compare the execution time versus PSNR with two state-of-the-art approaches with different reconstruction schemes. The first algorithm was proposed by Kalantari et al.~\cite{Kalantari2016Learning} which reconstructed the LF in an \emph{aperture-wise} manner. To generate each novel view, the algorithm must also estimate the disparity map and adjust the pixel color value, which further add to its overall run time. The second approach is an multi-step \emph{EPI-wise} reconstruction approach proposed by Wu et al.~\cite{Wu2018Light}, which called for blur-restoration-deblur steps for each EPI reconstruction. Fig.~\ref{fig:computations} shows the runtime versus PSNR performance over different schemes for $3\times3 \rightarrow 9\times9$ angular SR task. Compared with the EPI-wise~\cite{Wu2018Light} and aperture-wise~\cite{Kalantari2016Learning} method, our model is at least $40\times$ times faster.

\section{Conclusions}
\label{sec:conclusion}
In this paper, we proposed a high-dimensional deep convolutional network with dense connections for accurate LFSR. Our model progressively recovers spatio-angular information and high-frequency spatial details by minimizing MSE-based angular loss and content spatial loss. By introducing high-dimensional convolution layers, the proposed HDDRNet is able to reconstruct the light field at multiple scales in both spatial and angular dimensions. In addition, the extracted geometric features are sensitive to the object border and therefore indicate the scene geometric structure. To ease the training of such 4D framework, a novel normalization operation is defined based on a group of sub-aperture images in each feature map. Subsequently, we proposed the multi-range training strategy to further improve the reconstruction results, and named the improved model -- M-HDDRNet. Moreover, we also show the efficacy of the proposed M-HDDRNet in the context of recovering sub-pixel information in some challenging scenes.

\ifCLASSOPTIONcompsoc
  \section*{Acknowledgments}
  \label{sec:acknowledgements}
  This work was supported in part by the Hong Kong Research Grant Council (17203217 and 17201818) and the University of Hong Kong (104004142, 104004582, 104005009). We would like to thank the group of Dr. Wei-Ning Lee for their TiTan X GPU and server.
\else
  \section*{Acknowledgment}
  \label{sec:acknowledgements}
  This work was supported in part by the Hong Kong Research Grant Council (17203217 and 17201818) and the University of Hong Kong (104004142, 104004582, 104005009). We would like to thank the group of Dr. Wei-Ning Lee for their TiTan X GPU and server.
\fi

\ifCLASSOPTIONcaptionsoff
  \newpage
\fi



\bibliographystyle{IEEEtran}

\bibliography{IEEEabrv,bare_jrnl}

\begin{thebibliography}{10}
\providecommand{\url}[1]{#1}
\csname url@samestyle\endcsname
\providecommand{\newblock}{\relax}
\providecommand{\bibinfo}[2]{#2}
\providecommand{\BIBentrySTDinterwordspacing}{\spaceskip=0pt\relax}
\providecommand{\BIBentryALTinterwordstretchfactor}{4}
\providecommand{\BIBentryALTinterwordspacing}{\spaceskip=\fontdimen2\font plus
\BIBentryALTinterwordstretchfactor\fontdimen3\font minus
  \fontdimen4\font\relax}
\providecommand{\BIBforeignlanguage}[2]{{%
\expandafter\ifx\csname l@#1\endcsname\relax
\typeout{** WARNING: IEEEtran.bst: No hyphenation pattern has been}%
\typeout{** loaded for the language `#1'. Using the pattern for}%
\typeout{** the default language instead.}%
\else
\language=\csname l@#1\endcsname
\fi
#2}}
\providecommand{\BIBdecl}{\relax}
\BIBdecl

\bibitem{Ng2005Light}
R.~Ng, M.~Levoy, M.~Br{\'e}dif, G.~Duval, M.~Horowitz, and P.~Hanrahan, ``Light
  field photography with a hand-held plenoptic camera,'' \emph{Computer Science
  Technical Report}, vol.~2, no.~11, pp. 1--11, 2005.

\bibitem{Lam2015Computational}
E.~Y. Lam, ``Computational photography with plenoptic camera and light field
  capture: tutorial,'' \emph{Journal of the Optical Society of America A},
  vol.~32, no.~11, pp. 2021--2032, 2015.

\bibitem{Mitra2012Light}
K.~Mitra and A.~Veeraraghavan, ``Light field denoising, light field
  superresolution and stereo camera based refocussing using a {GMM} light field
  patch prior,'' in \emph{IEEE Conference on Computer Vision and Pattern
  Recognition Workshops}, July 2012, pp. 22--28.

\bibitem{Srinivasan2017Learning}
P.~P. Srinivasan, T.~Wang, A.~Sreelal, R.~Ramamoorthi, and R.~Ng, ``Learning to
  synthesize a 4{D} {RGBD} light field from a single image,'' in \emph{IEEE
  International Conference on Computer Vision}, vol.~2, no.~5, 2017, pp.
  2243--2251.

\bibitem{Kalantari2016Learning}
N.~K. Kalantari, T.-C. Wang, and R.~Ramamoorthi, ``Learning-based view
  synthesis for light field cameras,'' \emph{ACM Transactions on Graphics},
  vol.~35, no.~6, pp. 193:1--193:10, November 2016.

\bibitem{Wang2016Depth}
T.-C. Wang, A.~A. Efros, and R.~Ramamoorthi, ``Depth estimation with occlusion
  modeling using light-field cameras,'' \emph{IEEE Transactions on Pattern
  Analysis and Machine Intelligence}, vol.~38, no.~11, pp. 2170--2181, 2016.

\bibitem{Shin2018Epinet}
C.~Shin, H.-G. Jeon, Y.~Yoon, I.~S. Kweon, and S.~J. Kim, ``{EPINET}: A
  fully-convolutional neural network using epipolar geometry for depth from
  light field images,'' in \emph{IEEE Conference on Computer Vision and Pattern
  Recognition}, 2018, pp. 4748--4757.

\bibitem{Sun2016Data}
X.~Sun, Z.~Xu, N.~Meng, E.~Y. Lam, and H.~K.-H. So, ``Data-driven light field
  depth estimation using deep convolutional neural networks,'' in \emph{IEEE
  International Joint Conference on Neural Networks}, November 2016, pp.
  367--374.

\bibitem{Georgiev2010Focused}
T.~G. Georgiev and A.~Lumsdaine, ``Focused plenoptic camera and rendering,''
  \emph{Journal of Electronic Imaging}, vol.~19, no.~2, pp.
  021\,106--1--021\,106--11, April 2010.

\bibitem{Chan2007Super}
W.-S. Chan, E.~Y. Lam, M.~K. Ng, and G.~Y. Mak, ``Super-resolution
  reconstruction in a computational compound-eye imaging system,''
  \emph{Multidimensional Systems and Signal Processing}, vol.~18, no. 2-3, pp.
  83--101, February 2007.

\bibitem{Bishop2012Light}
T.~E. Bishop and P.~Favaro, ``The light field camera: Extended depth of field,
  aliasing, and superresolution,'' \emph{IEEE Transactions on Pattern Analysis
  and Machine Intelligence}, vol.~34, no.~5, pp. 972--986, August 2012.

\bibitem{Wanner2014Variational}
S.~Wanner and B.~Goldluecke, ``Variational light field analysis for disparity
  estimation and super-resolution,'' \emph{IEEE Transactions on Pattern
  Analysis and Machine Intelligence}, vol.~36, no.~3, pp. 606--619, August
  2014.

\bibitem{Yoon2015Learning}
Y.~Yoon, H.-G. Jeon, D.~Yoo, J.-Y. Lee, and I.~S. Kweon, ``Learning a deep
  convolutional network for light-field image super-resolution,'' in \emph{IEEE
  International Conference on Computer Vision Workshops}, February 2015, pp.
  57--65.

\bibitem{Wu2018Light}
G.~Wu, Y.~Liu, L.~Fang, Q.~Dai, and T.~Chai, ``Light field reconstruction using
  convolutional network on {EPI} and extended applications,'' \emph{IEEE
  Transactions on Pattern Analysis and Machine Intelligence}, no.~1, pp.
  1681--1694, 2018.

\bibitem{Ioffe2015Batch}
S.~Ioffe and C.~Szegedy, ``Batch normalization: Accelerating deep network
  training by reducing internal covariate shift,'' \emph{arXiv preprint
  arXiv:1502.03167}, 2015.

\bibitem{Johnson2016Perceptual}
J.~Johnson, A.~Alahi, and L.~Fei-Fei, ``Perceptual losses for real-time style
  transfer and super-resolution,'' in \emph{European Conference on Computer
  Vision}, vol. 9906, September 2016, pp. 694--711.

\bibitem{Liang2015Light}
C.-K. Liang and R.~Ramamoorthi, ``A light transport framework for lenslet light
  field cameras,'' \emph{ACM Transactions on Graphics}, vol.~34, no.~2, pp.
  16:1--16:19, February 2015.

\bibitem{Wu2017LightTIP}
G.~Wu, B.~Masia, A.~Jarabo, Y.~Zhang, L.~Wang, Q.~Dai, T.~Chai, and Y.~Liu,
  ``Light field image processing: An overview,'' \emph{IEEE Journal of Selected
  Topics in Signal Processing}, vol.~11, no.~7, pp. 926--954, August 2017.

\bibitem{Lim2009Improving}
J.~Lim, H.~Ok, B.~Park, J.~Kang, and S.~Lee, ``Improving the spatial resolution
  based on {4D} light field data,'' in \emph{IEEE International Conference on
  Image Processing}, February 2009, pp. 1173--1176.

\bibitem{Wang2018Lfnet}
Y.~Wang, F.~Liu, K.~Zhang, G.~Hou, Z.~Sun, and T.~Tan, ``{LFNet}: A novel
  bidirectional recurrent convolutional neural network for light-field image
  super-resolution,'' \emph{IEEE Transactions on Image Processing}, vol.~27,
  no.~9, pp. 4274--4286, 2018.

\bibitem{Farrugia2018Light}
R.~A. Farrugia and C.~Guillemot, ``Light field super-resolution using a
  low-rank prior and deep convolutional neural networks,'' \emph{IEEE
  Transactions on Pattern Analysis and Machine Intelligence}, 2019.

\bibitem{Jeon2015Accurate}
H.-G. Jeon, J.~Park, G.~Choe, J.~Park, Y.~Bok, Y.-W. Tai, and I.~So~Kweon,
  ``Accurate depth map estimation from a lenslet light field camera,'' in
  \emph{IEEE Conference on Computer Vision and Pattern Recognition}, October
  2015, pp. 1547--1555.

\bibitem{Tao2013Depth}
M.~W. Tao, S.~Hadap, J.~Malik, and R.~Ramamoorthi, ``Depth from combining
  defocus and correspondence using light-field cameras,'' in \emph{IEEE
  International Conference on Computer Vision}, March 2013, pp. 673--680.

\bibitem{Wang2015Occlusion}
T.-C. Wang, A.~A. Efros, and R.~Ramamoorthi, ``Occlusion-aware depth estimation
  using light-field cameras,'' in \emph{IEEE International Conference on
  Computer Vision}, February 2015, pp. 3487--3495.

\bibitem{Wanner2012Spatial}
S.~Wanner and B.~Goldluecke, ``Spatial and angular variational super-resolution
  of {4D} light fields,'' in \emph{European Conference on Computer
  Vision}.\hskip 1em plus 0.5em minus 0.4em\relax Springer, 2012, pp. 608--621.

\bibitem{Pearson2013Plenoptic}
J.~Pearson, M.~Brookes, and P.~L. Dragotti, ``Plenoptic layer-based modeling
  for image based rendering,'' \emph{IEEE Transactions on Image Processing},
  vol.~22, no.~9, pp. 3405--3419, June 2013.

\bibitem{Zhang2013Light}
Z.~Zhang, Y.~Liu, and Q.~Dai, ``Light field from micro-baseline image pair,''
  \emph{IEEE Conference on Computer Vision and Pattern Recognition}, pp.
  3800--3809, October 2015.

\bibitem{Zhang2017Plenopatch}
F.-L. Zhang, J.~Wang, E.~Shechtman, Z.-Y. Zhou, J.-X. Shi, and S.-M. Hu,
  ``{PlenoPatch}: Patch-based plenoptic image manipulation,'' \emph{IEEE
  Transactions on Visualization and Computer Graphics}, vol.~23, no.~5, pp.
  1561--1573, February 2017.

\bibitem{Levoy1996Light}
M.~Levoy and P.~Hanrahan, ``Light field rendering,'' in \emph{ACM Conference on
  Computer Graphics and Interactive Techniques}, 1996, pp. 31--42.

\bibitem{Levin2010Linear}
A.~Levin and F.~Durand, ``Linear view synthesis using a dimensionality gap
  light field prior,'' in \emph{IEEE Conference on Computer Vision and Pattern
  Recognition}, August 2010, pp. 1831--1838.

\bibitem{Vagharshakyan2018Light}
S.~Vagharshakyan, R.~Bregovic, and A.~Gotchev, ``Light field reconstruction
  using shearlet transform,'' \emph{IEEE Transactions on Pattern Analysis and
  Machine Intelligence}, vol.~40, no.~1, pp. 133--147, 2018.

\bibitem{Flynn2016Deepstereo}
J.~Flynn, I.~Neulander, J.~Philbin, and N.~Snavely, ``{DeepStereo}: Learning to
  predict new views from the world's imagery,'' in \emph{Proceedings of the
  IEEE Conference on Computer Vision and Pattern Recognition}, December 2016,
  pp. 5515--5524.

\bibitem{Gul2018Spatial}
M.~S.~K. Gul and B.~K. Gunturk, ``Spatial and angular resolution enhancement of
  light fields using convolutional neural networks,'' \emph{IEEE Transactions
  on Image Processing}, vol.~27, no.~5, pp. 2146--2159, May 2018.

\bibitem{Yoon2017Light}
Y.~Yoon, H.-G. Jeon, D.~Yoo, J.-Y. Lee, and I.~S. Kweon, ``Light-field image
  super-resolution using convolutional neural network,'' \emph{Signal
  Processing Letters}, vol.~24, no.~6, pp. 848--852, 2017.

\bibitem{Gortler1996Lumigraph}
S.~J. Gortler, R.~Grzeszczuk, R.~Szeliski, and M.~F. Cohen, ``The lumigraph,''
  in \emph{ACM Annual Conference on Computer Graphics and Interactive
  Techniques}, 1996, pp. 43--54.

\bibitem{Wu2017Light}
G.~Wu, M.~Zhao, L.~Wang, Q.~Dai, T.~Chai, and Y.~Liu, ``Light field
  reconstruction using deep convolutional network on {EPI},'' in \emph{IEEE
  Conference on Computer Vision and Pattern Recognition}, vol. 2017, November
  2017, pp. 1638--1646.

\bibitem{Maas2013Rectifier}
A.~L. Maas, A.~Y. Hannun, and A.~Y. Ng, ``Rectifier nonlinearities improve
  neural network acoustic models,'' in \emph{International Conference on
  Machine Learning}, vol.~30, no.~1, 2013.

\bibitem{Shi2016Real}
W.~Shi, J.~Caballero, F.~Husz{\'a}r, J.~Totz, A.~P. Aitken, R.~Bishop,
  D.~Rueckert, and Z.~Wang, ``Real-time single image and video super-resolution
  using an efficient sub-pixel convolutional neural network,'' in \emph{IEEE
  Conference on Computer Vision and Pattern Recognition}, June 2016, pp.
  1874--1883.

\bibitem{Ledig2016Photo}
C.~Ledig, L.~Theis, F.~Husz{\'a}r, J.~Caballero, A.~Cunningham, A.~Acosta,
  A.~Aitken, A.~Tejani, J.~Totz, Z.~Wang, and W.~Shi, ``Photo-realistic single
  image super-resolution using a generative adversarial network,'' \emph{IEEE
  Conference on Computer Vision and Pattern Recognition}, pp. 105--114,
  November 2017.

\bibitem{Gupta2011Modified}
P.~Gupta, P.~Srivastava, S.~Bhardwaj, and V.~Bhateja, ``A modified {PSNR}
  metric based on {HVS} for quality assessment of color images,'' in \emph{IEEE
  International Conference on Communication and Industrial Application},
  February 2011, pp. 1--4.

\bibitem{Simonyan2014Very}
K.~Simonyan and A.~Zisserman, ``Very deep convolutional networks for
  large-scale image recognition,'' \emph{International Conference on Learning
  Representations}, 2015.

\bibitem{Glorot2010Understanding}
X.~Glorot and Y.~Bengio, ``Understanding the difficulty of training deep
  feedforward neural networks,'' in \emph{The thirteenth International
  Conference on Artificial Intelligence and Statistics}, 2010, pp. 249--256.

\bibitem{Gross2016Training}
G.~Sam and W.~Michael, ``Training and investigating residual nets,''
  \url{http://torch.ch/blog/2016/02/04/resnets.html}, February 2016.

\bibitem{Lai2017Fast}
W.-S. Lai, J.-B. Huang, N.~Ahuja, and M.-H. Yang, ``Fast and accurate image
  super-resolution with deep laplacian pyramid networks,'' \emph{arXiv preprint
  arXiv:1710.01992}, 2017.

\bibitem{Farrugia2017Super}
R.~A. Farrugia, C.~Galea, and C.~Guillemot, ``Super resolution of light field
  images using linear subspace projection of patch-volumes,'' \emph{IEEE
  Journal of Selected Topics in Signal Processing}, vol.~11, no.~7, pp.
  1058--1071, August 2017.

\bibitem{Cheng2019Light}
Z.~Cheng, Z.~Xiong, C.~Chen, and D.~Liu, ``Light field super-resolution: a
  benchmark,'' in \emph{IEEE Conference on Computer Vision and Pattern
  Recognition}, October 2019.

\bibitem{Rossi2018Geometry}
M.~Rossi and P.~Frossard, ``Geometry-consistent light field super-resolution
  via graph-based regularization,'' \emph{IEEE Transactions on Image
  Processing}, vol.~27, no.~9, pp. 4207--4218, 2018.

\bibitem{StanfordLytro}
``Stanford {Lytro} light field archive,''
  \url{http://lightfields.stanford.edu/}.

\bibitem{Ziegler2017Acquisition}
M.~Ziegler, R.~op~het Veld, J.~Keinert, and F.~Zilly, ``Acquisition system for
  dense lightfield of large scenes,'' in \emph{IEEE Conference on The True
  Vision-Capture, Transmission and Display of 3D Video}, February 2017, pp.
  1--4.

\bibitem{Abadi2016Tensorflow}
M.~Abadi, P.~Barham, J.~Chen, Z.~Chen, A.~Davis, J.~Dean, M.~Devin,
  S.~Ghemawat, G.~Irving, M.~Isard, M.~Kudlur, J.~Levenberg, R.~Monga,
  S.~Moore, D.~G. Murray, B.~Steiner, P.~Tucker, V.~Vasudevan, P.~Warden,
  M.~Wicke, Y.~Yu, and X.~Zheng, ``{TensorFlow}: A system for large-scale
  machine learning,'' in \emph{The 12th {USENIX} Symposium on Operating Systems
  Design and Implementation}, Savannah, GA, 2016, pp. 265--283.

\bibitem{Rerabek2016New}
M.~Rerabek and T.~Ebrahimi, ``New light field image dataset,'' in \emph{The 8th
  International Conference on Quality of Multimedia Experience}, no.
  EPFL-CONF-218363, June 2016.

\bibitem{Mousnier2015Partial}
A.~Mousnier, E.~Vural, and C.~Guillemot, ``Partial light field tomographic
  reconstruction from a fixed-camera focal stack,'' \emph{arXiv preprint
  arXiv:1503.01903}, 2015.

\bibitem{Honauer2016Dataset}
K.~Honauer, O.~Johannsen, D.~Kondermann, and B.~Goldluecke, ``A dataset and
  evaluation methodology for depth estimation on 4{D} light fields,'' in
  \emph{Asian Conference on Computer Vision}.\hskip 1em plus 0.5em minus
  0.4em\relax Springer, March 2016, pp. 19--34.

\bibitem{Wanner2013Datasets}
S.~Wanner, S.~Meister, and B.~Goldluecke, ``Datasets and benchmarks for densely
  sampled {4D} light fields.'' in \emph{Vision, Modeling, and Visualization},
  vol.~13, 2013, pp. 225--226.

\bibitem{Levoy2006Light}
M.~Levoy, R.~Ng, A.~Adams, M.~Footer, and M.~Horowitz, ``Light field
  microscopy,'' in \emph{ACM Transactions on Graphics}, vol.~25, no.~3, July
  2006, pp. 924--934.

\bibitem{Lin2015Camera}
X.~Lin, J.~Wu, G.~Zheng, and Q.~Dai, ``Camera array based light field
  microscopy,'' \emph{Biomedical Optics Express}, vol.~6, no.~9, pp.
  3179--3189, 2015.

\bibitem{Wilburn2005High}
B.~Wilburn, N.~Joshi, V.~Vaish, E.-V. Talvala, E.~Antunez, A.~Barth, A.~Adams,
  M.~Horowitz, and M.~Levoy, ``High performance imaging using large camera
  arrays,'' in \emph{ACM Transactions on Graphics}, vol.~24, no.~3, 2005, pp.
  765--776.

\bibitem{Huang2017Densely}
G.~Huang, Z.~Liu, L.~Van Der~Maaten, and K.~Q. Weinberger, ``Densely connected
  convolutional networks.'' in \emph{IEEE Conference on Computer Vision and
  Pattern Recognition}, vol.~1, no.~2, November 2017, pp. 4700--4708.

\bibitem{Kim2016Accurate}
J.~Kim, J.~Kwon~Lee, and K.~Mu~Lee, ``Accurate image super-resolution using
  very deep convolutional networks,'' in \emph{IEEE Conference on Computer
  Vision and Pattern Recognition}, June 2016, pp. 1646--1654.

\bibitem{Zhang2018Residual}
Y.~Zhang, Y.~Tian, Y.~Kong, B.~Zhong, and Y.~Fu, ``Residual dense network for
  image super-resolution,'' in \emph{IEEE Conference on Computer Vision and
  Pattern Recognition}, 2018, pp. 2472--2481.

\bibitem{Jo2018Deep}
Y.~Jo, S.~W. Oh, J.~Kang, and S.~J. Kim, ``Deep video super-resolution network
  using dynamic upsampling filters without explicit motion compensation,'' in
  \emph{IEEE Conference on Computer Vision and Pattern Recognition}, 2018, pp.
  3224--3232.

\bibitem{Yeung2018Fast}
H.~W.~F. Yeung, J.~Hou, J.~Chen, Y.~Y. Chung, and X.~Chen, ``Fast light field
  reconstruction with deep coarse-to-fine modeling of spatial-angular clues,''
  in \emph{The European Conference on Computer Vision}, September 2018.

\end{thebibliography}
\end{document}